\documentclass[journal]{IEEEtran}
\usepackage{amsmath,amsfonts,amssymb,mathtools,dsfont,multicol,multirow,hhline,diagbox,dsfont,mathtools}
\usepackage{graphicx,hhline}
\usepackage{algorithmic,algorithm,threeparttable,url,subfigure,booktabs}
\usepackage{amsthm,color,bigstrut,capt-of}
\usepackage[noadjust]{cite}

\newtheorem{lemma}{Lemma}
\newtheorem{prop}{Proposition}

\newcommand{\Fopt}{\mathbf{F}_\mathrm{opt}}
\newcommand{\FRF}{\mathbf{F}_\mathrm{RF}}

\newcommand{\FBB}{\mathbf{F}_\mathrm{BB}}
\newcommand{\FDD}{\mathbf{F}_\mathrm{DD}}
\newcommand{\FBD}{\mathbf{F}_\mathrm{BD}}

\newcommand{\WRF}{\mathbf{W}_\mathrm{RF}}
\newcommand{\NRFr}{N_\mathrm{RF}^\mathrm{r}}
\newcommand{\NRFt}{N_\mathrm{RF}^\mathrm{t}}
\newcommand{\Nt}{N_\mathrm{t}}
\newcommand{\Nr}{N_\mathrm{r}}
\DeclareMathOperator{\Tr}{tr}

\newcounter{longequ}[longequ]
\addtocounter{longequ}{35}

\ifCLASSINFOpdf
\else
\fi

\begin{document}
\title{A Hardware-Efficient Analog Network Structure for Hybrid Precoding in Millimeter Wave Systems}

\author{Xianghao~Yu,~\IEEEmembership{Student Member,~IEEE},
        Jun~Zhang,~\IEEEmembership{Senior Member,~IEEE},
        and~Khaled~B.~Letaief,~\IEEEmembership{Fellow,~IEEE}
\thanks{This work was supported in part by the Hong Kong Research
	Grants Council under Grant No. 16210216. This paper was presented in part at the IEEE  Global Communications Conference (GLOBECOM), Singapore, Dec. 2017 \cite{gc}.
	
	X. Yu, J. Zhang and K. B. Letaief are with the Department of Electronic and Computer
	Engineering, the Hong Kong University of Science and Technology (HKUST),
	Kowloon, Hong Kong (e-mail: {xyuam, eejzhang,eekhaled}@ust.hk).
	
	K. B. Letaief is also with Hamad Bin Khalifa University, Doha, Qatar (e-mail: kletaief@hbku.edu.qa).
}
}

\maketitle

\begin{abstract}
	Hybrid precoding has been recently proposed as a cost-effective transceiver solution for millimeter wave (mm-wave) systems. 
	While the number of radio frequency (RF) chains has been effectively reduced in existing works, a large number of high-precision phase shifters are still needed. Practical phase shifters are with coarsely quantized phases, and their number should be reduced to a minimum due to cost and power consideration.
	In this paper, we propose a novel hardware-efficient implementation for hybrid precoding, called the \emph{fixed phase shifter} (FPS) implementation. It only requires a small number of phase shifters with quantized and \emph{fixed} phases.
	To enhance the spectral efficiency, a switch network is put forward to provide dynamic connections from phase shifters to antennas, which is adaptive to the channel states.
	An effective alternating minimization (AltMin) algorithm is developed with closed-form solutions in each iteration to determine the hybrid precoder and the states of switches. 
	Moreover, to further reduce the hardware complexity, a \emph{group-connected} mapping strategy is proposed to reduce the number of switches.
	Simulation results show that the FPS fully-connected hybrid precoder achieves higher hardware efficiency  with much fewer phase shifters than existing proposals. Furthermore, the group-connected mapping achieves a good balance between spectral efficiency and hardware complexity. 
\end{abstract}

\IEEEpeerreviewmaketitle

\section{Introduction}
Uplifting the carrier frequency to millimeter wave (mm-wave) bands is an effective approach to meet the capacity requirement of the upcoming 5G networks, and thus mm-wave communication has drawn extensive attention from both academia and industry \cite{6515173,6824752}. Thanks to the small wavelength of mm-wave signals, large-scale antenna arrays can be leveraged at transceivers to combat huge path loss at mm-wave frequencies and support directional transmissions with advanced multiple-input-multiple-output (MIMO) techniques. As equipping each antenna element with a single radio frequency (RF) chain is costly and power hungry, hybrid precoding has been put forward as a cost-effective transceiver solution, which utilizes a limited number of RF chains to connect a digital baseband precoder and an analog RF precoder \cite{6717211}.

In contrast to the conventional fully digital precoder, the additional hardware in the hybrid one is the analog component, also called the \emph{analog network}, which determines the overall hardware structure of the hybrid precoder. Most existing works on hybrid precoding are performance-oriented, i.e., aiming at maximizing the spectral efficiency \cite{7389996,6717211,7335586}. However, spectral efficiency close to the fully digital precoder was achieved with bulky hardware and impractical assumptions for the analog network, which results in a poor hardware efficiency and hinders its practical implementation.
Thus, it is of great importance to develop hardware-efficient analog networks that help the practical deployment of hybrid precoders.

To discuss hardware-efficient design, we first introduce a few terminologies for describing the hybrid precoder structure.
Each hybrid precoder structure is specified by its \emph{mapping strategy} and \emph{hardware implementation}. 
Specifically, the \emph{mapping strategy} decides how the RF chains and antenna elements are connected, which also determines the number of hardware components needed in the analog network.
Typical mapping strategies include the fully- and partially-connected ones. 
The fully-connected one exploits all the degrees of freedom to perform the mapping, i.e., it maps every RF chain to all the antennas, e.g., \cite{6717211}. In contrast, each RF chain is only connected to a subset of antennas in the partially-connected one, e.g., \cite{7010533}. 
On the other hand, the \emph{hardware implementation} specifies the adopted hardware components and the way each RF chain-antenna pair is connected.
The single phase shifter (SPS) implementation is the most commonly adopted one, which deploys one phase shifter to realize each RF chain-antenna connection \cite{1519678}. More recently, a double phase shifter (DPS) implementation was proposed in \cite{asilomar,spawc} to simplify the hybrid precoding algorithm design, where two distinct phase shifters are used to connect each RF chain-antenna pair.

In this paper, we propose a novel analog network structure that significantly improves the hardware efficiency of hybrid precoders. This is achieved by an innovative hardware implementation, called the \emph{fixed phase shifter} (FPS) implementation, and a new mapping strategy, i.e., the \emph{group-connected} mapping. In particular, the new structure can approach the performance of the fully digital precoder with very few fixed phase shifters.

\subsection{Related Works}
The fully-connected mapping strategy with the SPS implementation, referred as the SPS fully-connected structure, is the most popular structure in earlier works on hybrid precoding \cite{6717211,6928432,7335586,7227015,7037444}. However, this structure entails a drawback in the analog network, i.e., the number of phase shifters in use is $\NRFt\Nt$, with $\NRFt$ and $\Nt$ being the numbers of RF chains and antennas, respectively. 
Note that phase shifters, originally utilized in military radar systems, are newly-introduced hardware components in hybrid precoding systems, and currently very costly for commercial use, e.g., it can be around a hundred US dollars even with low resolution \cite{web}.
Hence, deploying such a large number of phase shifters would cause prohibitively high cost and power consumption. More importantly, phase shifters are assumed with \emph{variable high resolution} to provide near-optimal performance with effective algorithms, which is far from practical. 

To improve the hardware efficiency, one possible way is to reduce the number of phase shifters in use via changing the mapping strategy. Partially-connected mapping, which connects each RF chain to a subset of antennas, stands out as a popular solution \cite{7397861,7445130,7010533,7880698,spawc}. A semidefinite relaxation based alternating minimization (SDR-AltMin) algorithm was proposed in \cite{7397861} for hybrid precoder design with this mapping strategy. Based on a similar idea as successive interference cancellation (SIC), an iterative hybrid precoding algorithm for the partially-connected mapping was proposed in \cite{7445130}. In addition, a greedy algorithm and a modified K-means algorithm were developed in \cite{7880698} and \cite{spawc}, respectively, to dynamically optimize the subarrays in the partially-connected mapping for performance improvement. While various techniques were introduced to design hybrid precoders with the partially-connected mapping, there still exists a non-negligible gap in spectral efficiency compared with the fully-connected one. Inevitably, trade-offs need to be made between hardware efficiency and spectral efficiency, but the partially-connected mapping goes to an extreme, i.e., it enhances the hardware efficiency by incurring too much performance degradation. It is thus of practical importance to develop hardware-efficient hybrid precoder structures that can achieve more flexible trade-offs.

On the other hand, different hybrid precoding algorithms have been proposed assuming phase shifters with arbitrary precision, e.g., orthogonal matching pursuit (OMP) \cite{6717211}, manifold optimization \cite{7397861}, and SIC \cite{7445130}. 
Following these works, a straightforward refinement for practical hardware implementation is to design hybrid precoders with \emph{quantized} phase shifters \cite{7227015,7827111,6928432,7387790,7858800}. 
The main approach is either to determine all the phases at once \cite{6717211,7827111,6928432,7387790} or update one phase at a time \cite{7858800} by ignoring the quantization effect at first. Then the phases are heuristically quantized into the finite feasible set according to certain criteria. However, a simple quantization step is far from satisfactory, and the optimality and convergence of the proposed algorithms cannot be guaranteed \cite{7858800}. 
In addition, hybrid precoder design based on codebooks consisting of quantized phases was investigated in \cite{6824962,7448873,7006720}. While codebook-based design enjoys a low complexity, there will be certain performance loss, and it is not clear how much performance gain can be further obtained. 
The number of quantized phase shifters was to some extent reduced in \cite{7387790}, which is approximately $\lceil80p\rceil$ for achieving a certain required precision $\epsilon=10^{-p}$, 
e.g., around $160$ quantized phase shifters are needed for $\epsilon=0.01$.
Unfortunately, a large number of phase shifters are still needed for achieving a high spectral efficiency under practical settings in multiuser OFDM systems, i.e., $\sim$40 quantized phase shifters for each RF chain, and the number varies with the precision requirement.
More importantly, in these existing works, the phases need to be adapted to the channel states, which brings high hardware implementation complexity and also increases power consumption. Recently, a hybrid precoder structure that adopts switches to improve the hardware efficiency was put forward in \cite{7370753}. Nevertheless, simply replacing variable phase shifters with switches will cause significant performance degradation. Therefore, a more effective approach to handle quantized phases is needed, and the number of phase shifters should be reduced to a minimum.

\begin{figure*}
	\centering
	\includegraphics[height=4.7cm]{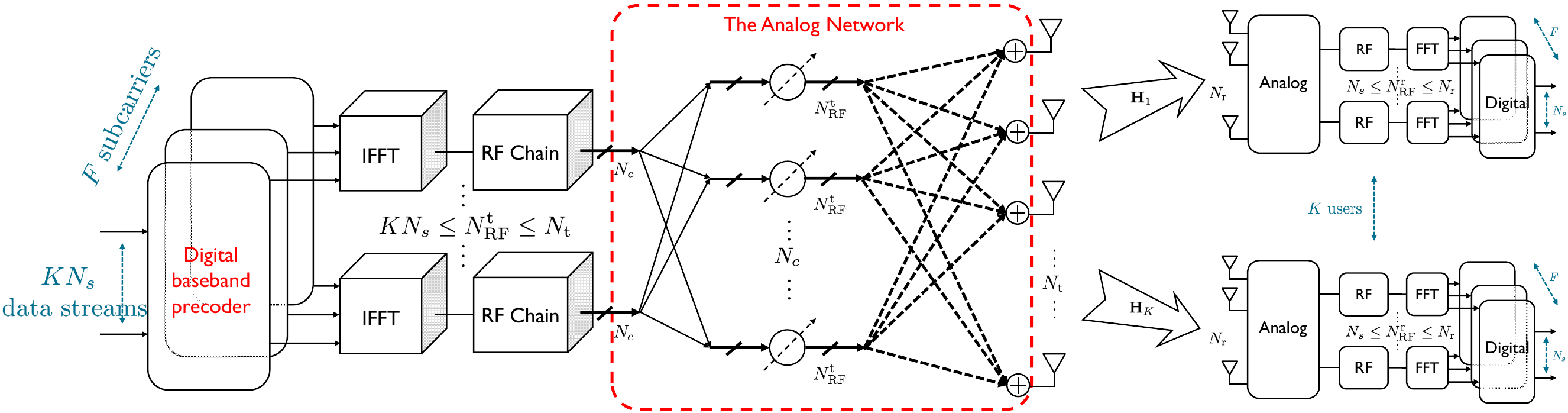}
	\caption{A multiuser mm-wave MIMO-OFDM system with FPS hybrid precoder implementation. To simplify the figure, in the analog precoder, each solid line with a slash represents parallel signal transmissions while each dotted line stands for $\NRFt$ switches. 
	}
	\label{systemmodel}
\end{figure*}

\subsection{Contributions}
In this paper, we investigate hardware-efficient design for hybrid precoding in general multiuser orthogonal frequency-division multiplexing (OFDM) mm-wave systems. The main contributions are summarized as follows.
\begin{itemize}
	\item As a first step, a novel hardware implementation is proposed for the analog network, called the fixed phase shifter (FPS) implementation, where only a small number of phase shifters with \emph{fixed} phases are needed. To compensate the performance loss induced by the fixed phases, a switch network is proposed to provide dynamic connections from phase shifters to antennas, which is easily implementable by adaptive switches.
	\item An AltMin algorithm is developed to design the hybrid precoder with the fully-connected mapping, where an upper bound of the objective function is derived as an effective surrogate. In particular, the large-scale binary constraints induced by the switch network are delicately tackled with the help of the upper bound, which leads to closed-from solutions for both the dynamic switch network and the digital baseband precoder, and therefore enables a low-complexity hybrid precoding algorithm.
	\item To further reduce the hardware complexity, a novel mapping strategy, i.e., the \emph{group-connected} mapping, is proposed and then applied along with the FPS implementation. This flexible mapping strategy incorporates the popular fully- and partially-connected mapping strategies as special cases. More importantly, the introduction of this new mapping strategy does not incur any additional design challenges as the hybrid precoder can be readily designed by leveraging existing hybrid precoding algorithms.
	\item Extensive comparisons are provided to reveal valuable design insights. 
	In particular, the FPS fully-connected hybrid precoder structure is shown to be able to easily approach the performance of the fully digital precoder, and enjoys a higher hardware efficiency than existing proposals. What deserves a special mention is the sharp reduction of the number of phase shifters compared with existing hybrid precoder implementations, e.g., $\sim$10 fixed phase shifters in total are sufficient. In addition, the FPS group-connected structure, which further reduces the number of switches, provides a flexible way to trade off spectral efficiency with hardware complexity.
\end{itemize}

In summary, our results firmly show that the proposed FPS group-connected structure is a promising candidate for hardware-efficient hybrid precoding in 5G mm-wave communication systems.

\subsection{Organization}
The remainder of this paper is organized as follows. In Section \ref{SecII}, we introduce the system model and proposed FPS implementation, followed by the problem formulation. The AltMin algorithms for the single-carrier and multicarrier systems with the FPS fully-connected mapping strategy are demonstrated in Sections \ref{SecIII} and \ref{SecIV}, respectively. Section \ref{SecV} introduces the group-connected mapping strategy. Simulation results are presented in Section \ref{SecVI}. Finally, we conclude this paper in Section \ref{SecVII}.

\subsection{Notations}
The following notations are used throughout this paper. $\mathbf{a}$ and $\mathbf{A}$ stand for a column vector and a matrix, respectively; 
The conjugate, transpose, and conjugate transpose of $\mathbf{A}$ are represented by $\mathbf{A}^*$, $\mathbf{A}^T$, and $\mathbf{A}^H$; $\left\|\mathbf{a}\right\|_2$ and $\left\|\mathbf{A}\right\|_F$ denote the $\ell_2$ and Frobenius norms of vector $\mathbf{a}$ and matrix $\mathbf{A}$; $\mathrm{blkdiag}(\mathbf{A}_1,\cdots,\mathbf{A}_i)$ establishes a block diagonal matrix using $\mathbf{A}_1,\cdots,\mathbf{A}_i$ as its diagonal terms;
$\mathrm{tr}(\mathbf{A})$ and $\mathrm{vec}(\mathbf{A})$ indicate the trace and vectorization; Expectation and the real part of a complex variable is noted by $\mathbb{E}[\cdot]$ and $\Re[\cdot]$. 

\section{System Model}\label{SecII}
\subsection{Hybrid Precoding and Combining}
Consider the downlink transmission of a multiuser mm-wave MIMO-OFDM system as shown in Fig. \ref{systemmodel}. A base station (BS) leverages an $\Nt$-size antenna array to serve $K$ users over $F$ subcarriers using OFDM. Each user is equipped with $\Nr$ antennas and receives $N_s$ data streams from the BS on each subcarrier. The numbers of available RF chains are $\NRFt$ and $\NRFr$ for the BS and each user, respectively, which are restricted as $KN_s\le\NRFt<\Nt$ and $N_s\le\NRFr<\Nr$.

The received signal of the $k$-th user on the $f$-th subcarrier is given by
\begin{equation}
\mathbf{y}_{k,f}=\sqrt{\rho_k}\mathbf{W}^H_{\mathrm{B}k,f}\mathbf{W}^H_{\mathrm{RF}k}\left(\mathbf{H}_{k,f}\FRF\sum_{k=1}^K{\mathbf{F}_\mathrm{B}}_{k,f}\mathbf{s}_{k,f}+\mathbf{n}_{k,f}\right),
\end{equation}
where the subscript $(k,f)$ stands for the $k$-th user on the $f$-th subcarrier. The average received power of the $k$-th user is denoted as $\rho_k$, and $\mathbf{s}_{k,f}$ is the transmitted signal such that $\mathbb{E}\left[\mathbf{s}_{k,f}\mathbf{s}_{k,f}^H\right]=\frac{P}{KN_sF}\mathbf{I}_{N_s}$, where $P$ is the transmit power. In addition, $\mathbf{n}_{k,f}$ denotes the circularly symmetric complex Gaussian noise with power as $\sigma_\mathrm{n}^2$ at the users. The digital baseband precoders and combiners are denoted as ${\mathbf{F}_\mathrm{B}}_{k,f}$ and ${\mathbf{W}_\mathrm{B}}_{k,f}$, respectively, with dimensions $\NRFt\times N_s$ and $\NRFr\times N_s$. Since the transmitted signals for all the users are mixed together by the digital precoders, and analog RF precoding is a post-IFFT (inverse fast Fourier transform) operation, the RF analog precoder $\FRF$ with dimension $\Nt\times\NRFt$ is a common component shared by all the users and subcarriers. Correspondingly, the $\Nr\times\NRFr$ RF analog combiner ${\WRF}_k$ is subcarrier-independent for each user. In this paper, we focus on the precoder design while the combiners can be designed in a similar way.

As discussed in Section I, each hybrid precoder structure is primarily determined by the \emph{mapping strategy} and \emph{hardware implementation}. In particular, the former maps the signals out of the limited RF chains to the large-scale antenna array, while the latter decides what kind of and how many hardware components are adopted to process the signal for each RF chain-antenna pair. In this section, a novel hardware implementation is first proposed to seek a hardware-efficient hybrid precoder structure. Then, to achieve a better balance between the hardware complexity and spectral efficiency, a flexible mapping strategy is introduced in Section \ref{SecV}.

\begin{table*}[htbp]
	\centering
	\caption{Comparisons of hardware components in the analog network for different hybrid precoder structures}
	\begin{tabular}{|c|l|c|c|c|c|c|c|}
		\hline
		\multicolumn{2}{|c|}{\multirow{3}[4]{*}{}} & \multicolumn{3}{c|}{\textbf{Phase shifter}} & \multicolumn{3}{c|}{\textbf{Other hardware components}} \bigstrut\\
		\cline{3-8}    \multicolumn{2}{|c|}{} & {\textbf{Number} {$N_\mathrm{PS}$}} & {\textbf{Type}} & \textbf{Power} {$P_\mathrm{PS}$} & {\textbf{Hardware}} & {\textbf{Number}} {$N_\mathrm{OC}$} & \textbf{Power} {$P_\mathrm{OC}$}\bigstrut[t]\\
		\hline
		\multirow{2}[4]{*}{\textbf{SPS \cite{6717211,7397861}}} & \textbf{Fully-connected} &  $\NRFt\Nt$     & \multirow{2}[4]{*}{Adaptive} & \multirow{2}[4]{*}{50 mW} & \multirow{2}[4]{*}{N/A} &    \multirow{2}[4]{*}{N/A}    & \multirow{2}[4]{*}{N/A} \bigstrut\\
		\cline{2-3}          & \textbf{Partially-connected} &   $\Nt$    &       &       &       &      &  \bigstrut\\
		\hline
		{\textbf{SPS with Butlter}} & \textbf{Fully-connected} &  $\frac{\NRFt\Nt}{2}(\log_2\Nt-1)$     & \multirow{2}[4]{*}{Fixed} & \multirow{2}[4]{*}{20 mW} & \multirow{2}[4]{*}{Coupler} & $\frac{\NRFt\Nt}{2}\log_2\Nt$      & \multirow{2}[4]{*}{10 mW} \bigstrut\\
		\cline{2-3}\cline{7-7}    \textbf{ matrices \cite{5345749}}     & \textbf{Partially-connected} &  $\frac{\Nt}{2}\left(\log_2\frac{\Nt}{\NRFt}-1\right)$     &       &       &       &  $\frac{\Nt}{2}\log_2\frac{\Nt}{\NRFt}$    &  \bigstrut\\
		\hline
		\multirow{2}[4]{*}{\textbf{DPS \cite{asilomar,spawc}}} & \textbf{Fully-connected} &  $2\NRFt\Nt$     & \multirow{2}[4]{*}{Adaptive} & \multirow{2}[4]{*}{50 mW} & \multirow{2}[4]{*}{N/A} &    \multirow{2}[4]{*}{N/A}   & \multirow{2}[4]{*}{N/A} \bigstrut\\
		\cline{2-3}       & \textbf{Partially-connected} & $2\Nt$      &       &       &       &       &  \bigstrut\\
		\hline
		\multirow{2}[4]{*}{\textbf{FPS}} & \textbf{Fully-connected} & \multirow{2}[4]{*}{$N_c\ll\Nt$}      & Multi-channel & \multirow{2}[4]{*}{20 mW} & \multirow{2}[4]{*}{Switch} &   $N_c\NRFt\Nt$    & \multirow{2}[4]{*}{5 mW} \bigstrut\\
		\cline{2-2}\cline{7-7}          & \textbf{Group-connected} &       & Fixed      &       &       & $\frac{1}{\eta}N_c\NRFt\Nt$       &  \bigstrut\\
		\hline
	\end{tabular}%
	\label{tab:addlabel}
\end{table*}%

\subsection{FPS Implementation}\label{SecIIB}
Recently, a DPS implementation was proposed in \cite{asilomar,spawc}, which enables low-complexity hybrid precoder design and also greatly improves the spectral efficiency.
These benefits come from allowing the same signal to pass through two phase shifters.
Inspired by this insight, we propose a hardware-efficient implementation in the following.

In the proposed implementation, $N_c$ phase shifters are used, where $N_c\ll\Nt$, as shown in Fig. \ref{systemmodel}. One critical difference between the proposed implementation and existing ones is that the number of phase shifters no longer depends on any other parameters, e.g., the number of RF chains or antennas, and can be made very small, which effectively improves the hardware efficiency. Inspired by the beneficial operation in the DPS implementation, the signal from each RF chain is passed through all $N_c$ available phase shifters. In other words, each phase shifter is an $\NRFt$-channel phase shifter \cite{6690170} that can simultaneously process the output signals from $\NRFt$ RF chains, i.e., in a parallel fashion.
On the other hand, while the number of (multi-channel) phase shifters could be small, it is still intractable to shift arbitrary phases or to switch between multiple quantized phase levels at a high speed to adapt to the channel states. In our proposal, instead of variable phase shifters, the $N_c$ phase shifters are assumed with \emph{fixed} phases \cite{5648370}, which is independent of the channel states. Thus, this proposal is referred as the FPS implementation.

\emph{Remark 1:} With the limited number of fixed phase shifters, the analog precoder can only provide the same static precoding gain for all RF chain-antenna pairs
and therefore inevitably entails performance loss. 

\begin{figure}[t]
	\centering
	\includegraphics[height=3cm]{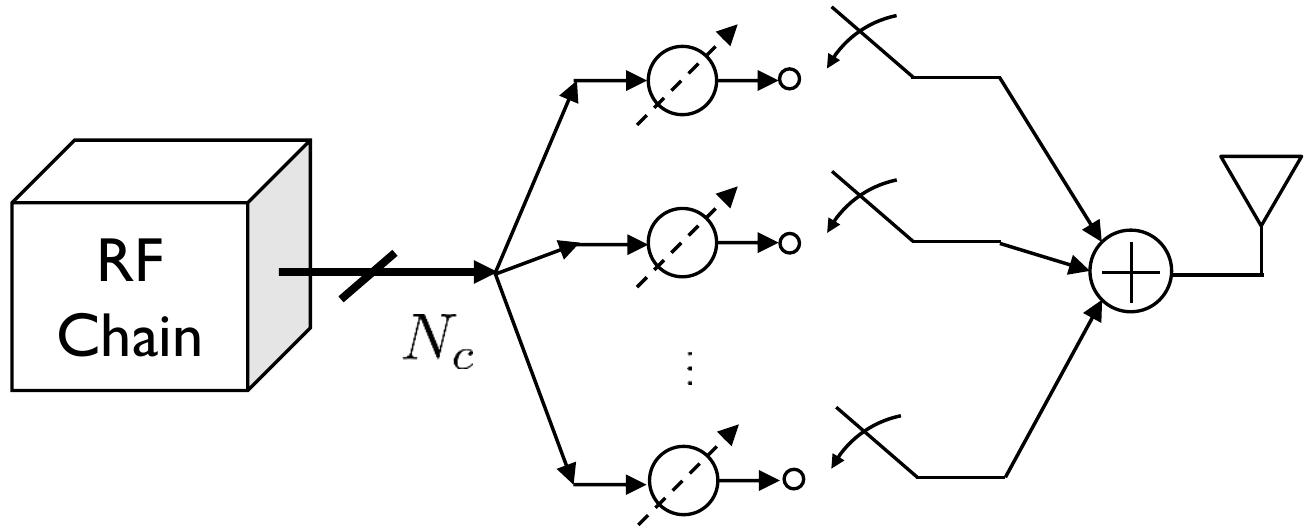}
	\caption{The FPS implementation from an RF chain to a connected antenna.
	}
	\label{sub}
\end{figure}

To overcome this drawback brought by the simplified hardware implementation, we propose to cascade a dynamic switch network after the fixed phase shifters, which is adapted to the channel states.
The signal flow in the FPS implementation is illustrated as follows.
To clearly illustrate the proposed FPS implementation, we focus on the signal flow of one RF chain-antenna pair, as shown in Fig. \ref{sub}.
The $N_c$ fixed phase shifters generate $N_c$ signals with different phases for the output signal of the given RF chain. We propose to adaptively combine a subset of the $N_c$ signals to compose the analog precoding gain from the RF chain to the antenna, which is realized by $N_c$ adaptive switches. Hence, $N_c$ switches are needed for each RF chain-antenna pair. Note that, with only binary on-off states, adaptive switches are much easier to implement than adaptive phase shifters \cite{5648370,7370753}. 

\emph{Remark 2:} The adaptive switch network enables the analog precoder to offer various precoding gains for different RF chain-antenna pairs to adapt to the channel states. Later we will see that although the proposed FPS implementation can only provide the analog precoding gains from a $\sum_{i=0}^{N_c}\binom{N_c}{i}$-dimension codebook, its performance is satisfactory with just a small value of $N_c$. 

In summary, all the hardware components needed for the FPS implementation are $N_c$ fixed phase shifters and $N_c$ switches per RF chain-antenna pair, and the total number of switches depends on the employed mapping strategy.

Accordingly, the analog RF precoding matrix $\FRF$ can be expressed as
\begin{equation}\label{SC}
\FRF=\mathbf{SC},
\end{equation}
where the switch matrix $\mathbf{S}$ is a binary matrix with dimension ${\Nt\times N_c\NRFt}$ , and the Boolean constraints are induced by the switches with binary states. Note that some entries may be forced to be zero due to different mapping strategies, which shall be discussed later. The matrix $\mathbf{C}\in\mathbb{C}^{N_c\NRFt\times\NRFt}$ stands for the phase shift operation carried out by the available fixed phase shifters, given by a block diagonal matrix as
\begin{equation}\label{psmatrix}
\mathbf{C}=\mathrm{blkdiag}\left(\underbrace{\mathbf{c},\mathbf{c},\cdots,\mathbf{c}}_{\NRFt}\right),
\end{equation}
where $\mathbf{c}=\frac{1}{\sqrt{N_c}}\left[e^{\jmath\theta_1},e^{\jmath\theta_2},\cdots,e^{\jmath\theta_{N_c}}\right]^T$ is the normalized phase shifter vector containing all $N_c$ fixed phases $\left\{\theta_i\right\}_{i=1}^{N_c}$. Note that although there are $N_c\NRFt$ non-zero parameters in matrix $\mathbf{C}$, only $N_c$ phase shifters are required since the phase shifters are with $\NRFt$ parallel channels and shared by all RF chain-antenna pairs.

Table \ref{tab:addlabel} lists the required hardware components in the analog network for different hybrid precoder structures, as well as the corresponding power consumption of each kind of hardware component \cite{7370753}. It shows that the proposed FPS implementation employs much less (fixed) phase shifters and consumes less power compared with existing works. While a bunch of switches are cascaded after the fixed phase shifters, the advantages of this proposal in hardware complexity and power consumption shall be demonstrated more explicitly in Section \ref{SecVI} via numerical comparisons.

\emph{Remark 3:} The ease of implementation and operation is another important aspect in hybrid precoder design. As switches only have binary states while high-resolution phase shifters need to be adaptive between a large number of states, the design and implementation of adaptive switches are generally easier than  high-resolution adaptive phase shifters \cite{belov2012handbook}, which makes the proposed FPS a practical and hardware-efficient implementation for the hybrid precoder structure.

\subsection{Problem Formulation}
There exist different formulations to maximize the spectral efficiency of hybrid precoding systems. One can either directly maximize the spectral efficiency \cite{7389996}, or adopt other performance metrics, e.g., mean square error (MSE) \cite{6874567} as surrogates to maximize the spectral efficiency. However, these formulations either result in high-complexity algorithms or with poor performance. More importantly, in multiuser multicarrier (MU-MC) systems, the analog precoder is a component that is shared by all users and subcarriers, which incurs additional difficulties on hybrid precoder design and therefore calls for a more tractable formulation to maximize the spectral efficiency.
It has been shown in \cite{6717211,7397861,7827111,7858800,asilomar,7037444,7914789} that minimizing the Euclidean distance between the fully digital precoder and the hybrid precoder is an effective and tractable alternative objective for maximizing the spectral efficiency in mm-wave systems.

On the other hand, it was found in \cite{asilomar,spawc} that the hybrid precoder in the multiuser setting produces residual inter-user interference, as it only approximates the fully digital precoder. Such interference will significantly degrade the system performance, especially at high SNR regimes. Moreover, this issue is more prominent in the multicarrier system as the analog precoder is shared by a large number of subcarriers. 

Therefore, to both effectively approximate the fully digital precoder and cancel the inter-user interference, we propose to apply a two-layer precoding at the baseband \cite{6542746}. In particular, the digital baseband precoder ${\mathbf{F}_\mathrm{B}}_{k,f}$ consists of two parts, i.e.,
\begin{equation}
{\mathbf{F}_\mathrm{B}}_{k,f}=\sqrt{\kappa}{\FBB}_{k,f}{\FBD}_{k,f},
\end{equation}
where $\kappa$ is a normalization factor, ${\FBB}_{k,f}\in\mathbb{C}^{\NRFt\times N_s}$ is the precoder that is utilized for approximating the fully digital precoder along with the analog precoder $\FRF$, and ${\FBD}_{k,f}\in\mathbb{C}^{N_s\times N_s}$ is the precoder that is responsible for canceling the inter-user interference. A similar approach was adopted in \cite{7954699}.

Correspondingly, the first task, i.e., to approximate the fully digital precoder, can be formulated as
\begin{equation}
\mathcal{P}_1:\quad
\begin{aligned}
&\underset{\mathbf{S},\mathbf{F}_\mathrm{BB}}{\mathrm{minimize}} && \left\Vert\Fopt-\mathbf{SC}\FBB\right\Vert_F^2\\
&\mathrm{subject\thinspace to}&&
\mathbf{S}\in\mathcal{B}
\end{aligned}
\end{equation}
where the combined fully digital precoder is denoted as $\Fopt=\left[{\Fopt}_{1,1},\cdots,{\Fopt}_{k,f},\cdots,{\Fopt}_{K,F}\right]\in\mathbb{C}^{\Nt\times KN_sF}$, and $\FBB=\left[{\FBB}_{1,1},\cdots,{\FBB}_{k,f},\cdots,{\FBB}_{K,F}\right]$ is the concatenated digital precoder\footnote{The phrase ``digital precoder'' is used to refer ${\FBB}_{k,f}$ in the remainder of this paper with a slight abuse of terminology, as it is the digital part in the hybrid precoder that approximates the fully digital precoder.} with dimension ${\NRFt\times KN_sF}$. 
The constraint set of the switch matrix is denoted as $\mathcal{B}$. 
Note that, while the transmit power constraint is not explicitly considered in $\mathcal{P}_1$, it shall be satisfied by adapting the normalization factor $\kappa$ after $\mathcal{P}_1$ is solved.

With the digital precoder ${\FBB}_{k,f}$ at hand, the other precoder ${\FBD}_{k,f}$ is cascaded after it to cancel the inter-user interference based on the effective channel including the hybrid precoder and physical channel, which is given by
\begin{equation}\label{effch}
\mathbf{\hat H}_{k,f}={\mathbf{W}^H_\mathrm{BB}}_{k,f}{\mathbf{W}^H_\mathrm{RF}}_{k}\mathbf{H}_{k,f}{\FRF}{\FBB}_{f},
\end{equation}
where ${\FBB}_f=\left[{\FBB}_{1,1},\cdots,{\FBB}_{k,f},\cdots,{\FBB}_{K,f}\right]$ with dimension ${\NRFt\times KN_s}$ is the composite digital precoder on the $f$-th subcarrier.
Then, our goal is to design precoders ${\FBD}_{k,f}$ that satisfy the conditions
\begin{equation}
\mathbf{\hat H}_{j,f}{\FBD}_{k,f}=\mathbf{0}, \quad k\ne j.
\end{equation}
A simple way to achieve the conditions is the block diagonal (BD) precoder. More details can be found in \cite{1261332}.

Since the inter-user interference is canceled, we can determine the normalization factor $\kappa$ to satisfy the transmit power constraint
$\sum_{k=1}^K\sum_{f=1}^F\left\Vert\FRF{\mathbf{F}_\mathrm{B}}_{k,f}\right\Vert_F^2\le KN_sF$, which is given by
\begin{equation}\label{kappa}
\kappa=\frac{KN_sF}{\sum_{k=1}^K\sum_{f=1}^F\left\Vert\mathbf{SC}{\FBB}_{k,f}{\FBD}_{k,f}\right\Vert_F^2}.
\end{equation}

Note that the combiners at the user side are with the same analog network structure as \eqref{SC}. The hybrid combiners can be designed in a similar way as $\mathcal{P}_1$ for each user independently, and thus are omitted due to space limitation.
In addition, the problem formulation is not limited to any specific channel models or fully digital precoding schemes. It can be easily observed that the hybrid precoder can be readily designed by \eqref{effch} to \eqref{kappa} once $\mathcal{P}_1$ is solved, and hence we will focus on $\mathcal{P}_1$ in the following sections.

\section{Hybrid Precoder Design With the FPS Implementation}\label{SecIII}
In this section, we design the hybrid precoder with the FPS implementation and the popular fully-connected mapping strategy, for which every entry in the switch matrix $\mathbf{S}$ is a binary optimization variable and there are in total $\Nt N_c\NRFt$ switches.
As shown in the hybrid precoder design problem $\mathcal{P}_1$, the main task is to design the binary switch matrix $\mathbf{S}$ and the digital precoding matrix $\FBB$. First we make some observations on $\mathcal{P}_1$.

\emph{Remark 4:} Since the switch matrix $\mathbf{S}$ is with finite possibilities, the cardinality of the constraint set $\mathcal{B}$ for the analog precoding matrix $\FRF$ is finite, which means that the OMP algorithm \cite{6717211,7037444} is applicable to $\mathcal{P}_1$. However, different from the SPS case, the dimension of the dictionary in the OMP algorithm for the FPS implementation is oversize, i.e., $\left[\sum_{i=0}^{N_c}\binom{N_c}{i}\right]^{\Nt}$, which is a huge number in large-scale antenna systems and hence hinders its practical implementation.

\emph{Remark 5:} Alternating minimization can be directly applied to $\mathcal{P}_1$ where the binary constraints can be tackled with the semidefinite relaxation (SDR) technique \cite{7397861}. However, an $\Nt\NRFt N_c+1$-dimension semidefinite programming (SDP) problem should be solved in each iteration, which causes prohibitive computational complexity. Moreover, how to recover a rank-one solution from an SDR with binary constraints is still an open problem \cite{luo2010semidefinite}. This means that the optimality of the relaxation in each iteration of the alternating procedure cannot be ensured and hence the overall convergence of the AltMin algorithm cannot be guaranteed.

As discussed above, the main difficulty to solve $\mathcal{P}_1$ is the large-size binary constraints of the switch matrix $\mathbf{S}$. As a matter of fact, even if we only focus on the design of the switch matrix $\mathbf{S}$, $\mathcal{P}_1$ is an NP-hard problem \cite{luo2010semidefinite}. In this section, by deriving an effective surrogate for the objective function and adopting alternating minimization, we come up with a low-complexity hybrid precoding algorithm that well tackles the binary constraints.

Note that the property of the combined digital precoding matrix $\FBB\in\mathbb{C}^{\NRFt\times KN_sF}$ differs for different system settings. It is a tall matrix in single-carrier systems, i.e., $F=1$, since $\NRFt\ge KN_s$. In contrast, when it comes to multicarrier systems, $\FBB$ is likely to be a fat matrix as $\NRFt<KN_sF$  for practical system parameters. As we will see in this section, this difference affects the manipulation of the algorithm, and we first present the hybrid precoder design in single-carrier systems\footnote{In this paper, single-carrier systems refer to single-carrier transmissions assuming flat-fading channels. The choice of such a model is for the ease of presentation, and the algorithm will be later extended to the more realistic multicarrier case with frequency-selective fading channels.}. 

\subsection{An Upper Bound for the Objective}
In \cite{7397861,asilomar,7389996}, it has been shown that imposing a semi-orthogonal structure for $\FBB$ is an efficient way to achieve near-optimal performance. Inspired by these results, we take a similar approach as follows. In single-carrier systems, the digital precoding matrix $\FBB$ is a tall matrix, and thus the semi-orthogonal constraint is specified as
\begin{equation}\label{eq5}
\FBB^H\FBB=\alpha^2\FDD^H\FDD=\alpha^2\mathbf{I}_{KN_s},
\end{equation}
where $\FBB=\alpha\FDD$, $\alpha$ is a scaling factor, and $\FDD$ is a semi-unitary matrix. 
Then, an upper bound is derived for the objective function in $\mathcal{P}_1$ in the following lemma.
\begin{lemma}
	The objective function in $\mathcal{P}_1$ is upper bounded by
	\begin{equation}\label{upperobj}
	\left\Vert\Fopt\right\Vert_F^2-2\alpha\Re\Tr\left(\FDD\Fopt^H\mathbf{SC}\right)+\alpha^2\left\Vert\mathbf{S}\right\Vert_F^2.
	\end{equation}
\end{lemma}
\begin{IEEEproof}
	The objective function in $\mathcal{P}_1$ can be rewritten as
	\begin{equation}\label{obj}
	\left\Vert\Fopt\right\Vert_F^2-2\alpha\Re\Tr\left(\FDD\Fopt^H\mathbf{SC}\right)+\alpha^2\left\Vert\mathbf{SC}\FDD\right\Vert_F^2.
	\end{equation}
	According to \eqref{psmatrix}, the phase shifter matrix $\mathbf{C}$ is a semi-unitary matrix, i.e., $\mathbf{C}^H\mathbf{C}=\mathbf{I}_{\NRFt}$. Therefore, we can derive an upper bound for the last term in \eqref{obj}, given by
	\begin{equation}\label{upper}
	\begin{split}
	\left\Vert\mathbf{SC}\FDD\right\Vert_F^2&=
	\Tr\left(\FDD^H\mathbf{C}^H\mathbf{S}^H\mathbf{SC}\FDD\right)\\
	&\overset{(a)}{=}\Tr \left( {
		\begin{bmatrix}
		\mathbf{I}_{KN_s}&\\
		&\mathbf{0}\\
		\end{bmatrix}
		{{\mathbf{K}^H}}{{\mathbf{S}^H}}{\mathbf{S}\mathbf{K}}} \right)\\
	&< \Tr\left( {{{\mathbf{K}}^H}{\mathbf{S}^H}}{\mathbf{S}\mathbf{K}} \right)=\left\Vert\mathbf{S}\right\Vert_F^2.
	\end{split}
	\end{equation}
	Step (a) follows the singular value decomposition (SVD) of $\mathbf{C}\FDD\FDD^H\mathbf{C}^H=\mathbf{K}\mathrm{blkdiag}\left(\mathbf{I}_{KN_s},\mathbf{0}\right)\mathbf{K}^H
	$ by utilizing the semi-unitary property of $\mathbf{C}\FDD$, whose left singular vectors are the columns of $\mathbf{K}$.
\end{IEEEproof}


\subsection{Alternating Minimization}\label{SecIIIB}
By adopting the upper bound \eqref{upperobj} as the surrogate objective function and dropping the constant term $\left\Vert\Fopt\right\Vert_F^2$, the hybrid precoder design problem $\mathcal{P}_1$ is reformulated as
\begin{equation}
\mathcal{P}_2:\quad
\begin{aligned}
&\underset{\alpha,\mathbf{S},\mathbf{F}_\mathrm{DD}}{\mathrm{minimize}} && \alpha^2\left\Vert\mathbf{S}\right\Vert_F^2-2\alpha\Re\Tr\left(\FDD\Fopt^H\mathbf{SC}\right)\\
&\mathrm{subject\thinspace to}&&
\begin{cases}
\mathbf{S}\in\{0,1\}^{\Nt\times N_c\NRFt}\\
\FDD^H\FDD=\mathbf{I}_{KN_s}.
\end{cases}
\end{aligned}
\end{equation}

Alternating minimization, as an effective tool for optimization problems involving different subsets of variables, has been widely applied and shown empirically successful in hybrid precoder design \cite{7397861,asilomar,7389996}. In this section, we apply this design principle to the hybrid precoder design with the FPS fully-connected structure.
In each step of the AltMin algorithm, one subset of the optimization variables is optimized while keeping the other parts fixed.

When the switch matrix $\mathbf{S}$ and $\alpha$ are fixed, the optimization problem can be written as
\begin{equation}
\begin{aligned}
&\underset{\mathbf{F}_\mathrm{DD}}{\mathrm{maximize}} && \alpha\Re\Tr\left(\FDD\Fopt^H\mathbf{SC}\right)\\
&\mathrm{subject\thinspace to}&&
\FDD^H\FDD=\mathbf{I}_{KN_s}.
\end{aligned}
\end{equation}
According to the definition of the dual norm \cite{horn2012matrix}, we have
\begin{equation}
\begin{split}
\alpha\Re\Tr\left(\FDD\Fopt^H\mathbf{SC}\right)&\le\left|\Tr\left(\alpha\FDD\Fopt^H\mathbf{SC}\right)\right|\\
&\overset{(b)}{\le}\left\Vert\FDD^H\right\Vert_\infty\left\Vert\alpha\Fopt^H\mathbf{SC}\right\Vert_1\\
&=\left\Vert\alpha\Fopt^H\mathbf{SC}\right\Vert_1=\sum_{i=1}^{KN_s}{\sigma_i},
\end{split}
\end{equation}
where $\left\Vert\cdot\right\Vert_\infty$ and $\left\Vert\cdot\right\Vert_1$ stand for the infinite and one Schatten norms \cite{horn2012matrix}, and (b) follows the H\"{o}lder's inequality. The equality is established only when
\begin{equation}\label{uv}
\FDD=\mathbf{V}_1\mathbf{U}^H,
\end{equation}
where $\alpha\Fopt^H\mathbf{SC}=\mathbf{U}\mathbf{\Sigma V}_1^H$ follows the SVD and $\mathbf{\Sigma}$ is a diagonal matrix with non-zero singular values $\sigma_1,\cdots,\sigma_{KN_s}$.

While we can divide the optimization of the two variables $\alpha$ and $\mathbf{S}$ into two separate subproblems, we propose to update them simultaneously to save the number of subproblems involved in the AltMin algorithm and therefore reduce the computational complexity. By adding a constant term $\left\Vert\Re\left(\Fopt\FDD^H\mathbf{C}^H\right)\right\Vert_F^2$ to the objective function in $\mathcal{P}_2$, the subproblem of updating $\alpha$ and $\mathbf{S}$ can be recast as
\begin{equation}\label{eq13}
\begin{aligned}
&\underset{\alpha,\mathbf{S}}{\mathrm{minimize}} && \left\Vert\Re\left(\Fopt\FDD^H\mathbf{C}^H\right)-\alpha\mathbf{S}\right\Vert_F^2\\
&\mathrm{subject\thinspace to}&&
\mathbf{S}\in\{0,1\}^{\Nt\times N_c\NRFt}.
\end{aligned}
\end{equation}
\begin{prop}
	The optimal solution to \eqref{eq13} is given by
	\begin{equation}\label{alpha}
	\alpha^\star=\arg\underset{\{\tilde{x}_i,{\bar{x}}_i\}_{i=1}^n}{\min}\quad\left\{f(2\tilde{x}_i),f({\bar{x}}_i)\right\},
	\end{equation}
	\begin{equation}\label{eq15}
	\mathbf{S}^\star=\begin{cases}
	\mathds{1}\left\{\Re\left(\Fopt\FDD^H\mathbf{C}^H\right)>\frac{\alpha}{2}\mathbf{1}_{\Nt\times N_c\NRFt}\right\}&\alpha>0\\
	\mathds{1}\left\{\Re\left(\Fopt\FDD^H\mathbf{C}^H\right)<\frac{\alpha}{2}\mathbf{1}_{\Nt\times N_c\NRFt}\right\}&\alpha<0,\\
	\end{cases}
	\end{equation}
	where $n=\Nt N_c\NRFt$, $\mathbf{x}=\mathrm{vec}\left\{\Re\left(\Fopt\FDD^H\mathbf{C}^H\right)\right\}$, $\mathds{1}(\cdot)$ is the indicator function, and $\mathbf{1}_{m\times n}$ denotes an $m\times n$ matrix with all entries equal to one. The objective function in \eqref{eq13} can be rewritten as $f(\alpha)$ in \eqref{longeq} in the proof. In addition, $\tilde{x}_i$ is the $i$-th smallest entry in $\mathbf{x}$, and
	\begin{equation}\label{eq16}
	{\bar{x}}_i\triangleq\begin{cases}
	\frac{\sum_{j=1}^i\tilde{x}_j}{i}&\bar{x}_i<0\text{ and }\bar{x}_i\in\mathcal{R}_i\\
	\frac{\sum_{j=i+1}^n\tilde{x}_j}{n-i}&\bar{x}_i>0\text{ and }\bar{x}_i\in\mathcal{R}_i\\
	+\infty&\text{otherwiese},
	\end{cases}
	\end{equation}
	where $\mathcal{R}_i\triangleq[2\tilde{x}_i,2\tilde{x}_{i+1}]$.
\end{prop}
\begin{IEEEproof}
	See Appendix A.
\end{IEEEproof}
Basically, $f(\alpha)$ is a quadratic function within each interval $\mathcal{R}_i$, as shown in \eqref{longeq} in the proof. This means that the optimal solutions of $\alpha$ in all the intervals $\{\mathcal{R}_i\}_{i=1}^n$ can only be obtained either at the endpoints of the intervals, i.e., $\{2\tilde{x}_i\}_{i=1}^n$, or at the vertexes of the parabolas, i.e., $\{\bar{x}_i\}_{i=1}^n$, if they fall into the intervals. Therefore, the optimal $\alpha^\star$ is obtained via a closed-form solution by comparing the optimal solutions of $\alpha$ in all the intervals $\{\mathcal{R}_i\}_{i=1}^n$, as indicated in \eqref{alpha}.
Nevertheless, since the number of intervals to be compared is $n=\Nt N_c\NRFt$, it will incur high computational complexity when $\Nt$ is large as in mm-wave systems. In the following lemma, we show that there is no need to compute the optimal $\alpha$ in all the intervals $\{\mathcal{R}_i\}_{i=1}^n$, which further reduces the complexity of the proposed algorithm.
\begin{lemma}\label{lem1}
	The optimal $\alpha^\star$ is obtained at one of the points ${\bar{x}}_i\in\mathcal{X}$, where $\mathcal{X}$ denotes the set of the ${\bar{x}}_i$'s that have finite values of $f({\bar{x}}_i)$.
\end{lemma}
\begin{IEEEproof}
	See Appendix B.
\end{IEEEproof}
Lemma \ref{lem1} indicates that any endpoints $\{2\tilde{x}_i\}_{i=1}^n$of the intervals $\{\mathcal{R}_i\}_{i=1}^n$ cannot be the optimal solution for $\alpha$. Moreover, since $f(\alpha)$ is a coercive function, i.e., $f(+\infty)\to+\infty$, we only need to pick the ${\bar{x}}_i$'s that have finite values of $f({\bar{x}}_i)$, i.e., the ones that satisfy the first two conditions in \eqref{eq16}, and the optimal solution for $\alpha$ is given by
\begin{equation}\label{eq20}
\alpha^\star=\arg\underset{{{\bar{x}}_i\in\mathcal{X}}}{\min}\quad f({\bar{x}}_i).
\end{equation}
By Lemma \ref{lem1}, the number of intervals we need to compare to obtain the optimal $\alpha^\star$ is shrunk from $n$ to $|\mathcal{X}|$, which is empirically shown to be less than 5 via simulations in Section \ref{SecVI} and hence further reduces the computational complexity of the proposed AltMin algorithm. 

Thus, we have shown that, with the help of the upper bound derived in \eqref{upper}, the large-scale binary switch matrix $\mathbf{S}$ can be efficiently optimized by a closed-form solution \eqref{eq15}, which verifies the benefits and superiority of the surrogate objective function adopted in $\mathcal{P}_2$.
With the closed-form solutions derived in \eqref{uv}, \eqref{eq15}, and \eqref{eq20} at hands, the AltMin algorithm for the FPS fully-connected structure in single-carrier systems is summarized as \textbf{FPS-AltMin Algorithm}. There are several issues involved in the FPS-AltMin algorithm that require some further remarks.

\floatname{algorithm}{FPS-AltMin Algorithm:}
\begin{algorithm}[t]
	\caption{A Low-Complexity Hybrid Precoding Algorithm for the FPS Fully-Connected Structure}
	\begin{algorithmic}[1]\label{lowcom}
		\REQUIRE
		$\Fopt$
		\STATE Construct an initial point for $\FDD^{(0)}$ according to \eqref{uv} and set $k=0$;
		\REPEAT 
		\STATE Fix $\FDD^{(k)}$, optimize $\alpha^{(k)}$ and $\mathbf{S}^{(k)}$ according to \eqref{eq20} and \eqref{eq15}, respectively;
		\STATE Fix $\mathbf{S}^{(k)}$ and $\alpha^{(k)}$, update $\FDD^{(k)}$ with \eqref{uv}; 
		\STATE $k\leftarrow k+1$;
		\UNTIL convergence.
		\STATE Compute the additional BD precoder ${\FBD}_{k,f}$ at the baseband to cancel the inter-user interference \cite{asilomar}, and calculate the normalization factor $\kappa$ according to \eqref{kappa} for the hybrid precoder at the transmit end.
		\RETURN $\FRF=\mathbf{SC}$ and ${\mathbf{F}_\mathrm{B}}_{k,f}=\alpha\sqrt{\kappa}{\FDD}_{k,f}{\FBD}_{k,f}$.
	\end{algorithmic}
\end{algorithm}

\emph{1) Convergence}: 
The FPS-AltMin algorithm is essentially a block coordinate descent (BCD) algorithm with two blocks $\FDD$ and $\{\mathbf{S},\alpha\}$, whose globally optimal solutions are given by \eqref{uv}, \eqref{eq15} and \eqref{eq20}. Hence, the algorithm is guaranteed to converge to a stationary point of $\mathcal{P}_2$ \cite{grippo2000convergence}. 

\emph{2) Initial point}: 
Since the algorithm converges to a stationary point, it may be sensitive to the initial point $\FDD^{(0)}$. We provide a way to construct an initial point in the FPS-AltMin algorithm.
The fully digital precoding matrix $\Fopt$ can be decomposed as follows according to its SVD $\Fopt=\mathbf{U\Sigma V}^H$, i.e.,
\begin{equation}\label{eq18}
\Fopt=\begin{bmatrix}
\mathbf{U\Sigma}&\mathbf{T}
\end{bmatrix}
\begin{bmatrix}
\mathbf{V}^H\\
\mathbf{0}
\end{bmatrix},
\end{equation} 
where $\mathbf{U\Sigma}$ is an $\Nt\times KN_s$ matrix with full column rank, $\mathbf{V}^H$ is a $KN_s$-dimension square matrix, and $\mathbf{T}$ is an arbitrary $\Nt\times(\NRFt-KN_s)$ matrix. In \eqref{eq18}, the fully digital precoding matrix $\Fopt$ is decomposed into two matrices that satisfy the dimensions of $\FRF$ and $\FDD$, respectively. In other words, $\FRF=\begin{bmatrix}
\mathbf{U\Sigma}&\mathbf{F}
\end{bmatrix}$, $\alpha=1$, and $\FDD=\begin{bmatrix}
\mathbf{V}&\mathbf{0}
\end{bmatrix}^H$ is a globally optimal solution to the hybrid precoding problem without any constraints on the analog precoding matrix $\FRF$. In this way, we generate the initial point $\FDD^{(0)}$ as
\begin{equation}\label{ini1}
\FDD^{(0)}=
\begin{bmatrix}
\mathbf{V}&\mathbf{0}_{KN_s\times(\NRFt-KN_s)}
\end{bmatrix}^H.
\end{equation}
Note that $\FDD^{(0)}$ fully extracts the information of the row space of $\Fopt$, whose basis are the first  $KN_s$ rows in $\FDD^{(0)}$. We also stress that the $\FDD^{(0)}$ satisfies the semi-unitary constraint introduced in \eqref{eq5}.

\emph{3) Computational complexity}:
We compare the computational complexity of the proposed algorithm with the ones mentioned in Remarks 4 and 5. Since the dictionary size in the OMP algorithm is $\left[\sum_{i=0}^{N_c}\binom{N_c}{i}\right]^{\Nt}$, the computational complexity could be prohibitively high even though this algorithm only needs a small number of iterations. For the SDR method mentioned in Remark 5, in each iteration\footnote{The procedure that updates both the analog and digital precoders is counted as one iteration.}, an $n+1$-dimension SDP problem should be solved for updating the analog part while a pseudo-inverse operation is needed for updating the digital precoder. Therefore, the computational complexity per iteration is $\mathcal{O}\left(\left(\Nt N_c\NRFt+1\right)^{6.5}\right)$. On the contrary, in each iteration of the proposed FPS-AltMin algorithm, the computational complexity is dominated by the truncated SVD and sorting operations, with the complexity $\mathcal{O}\left(K^2N_s^2\NRFt+N_c\NRFt\Nt\log N_c\NRFt\Nt\right)$, which is much lower than those of the OMP algorithm and SDR method\footnote{To solve the switch matrix $\mathbf{S}$ in one iteration, the running time of the SDR method is 1.3 s while the proposed FPS-AltMin algorithm takes 0.04 s when $\Nt=64$, $\Nr=16$, and $\NRFt=\NRFr=N_s=4$.}.

 \begin{figure*}[t]
	\centering
	\subfigure[Fully-connected mapping strategy.]{
		\includegraphics[height=5.4cm]{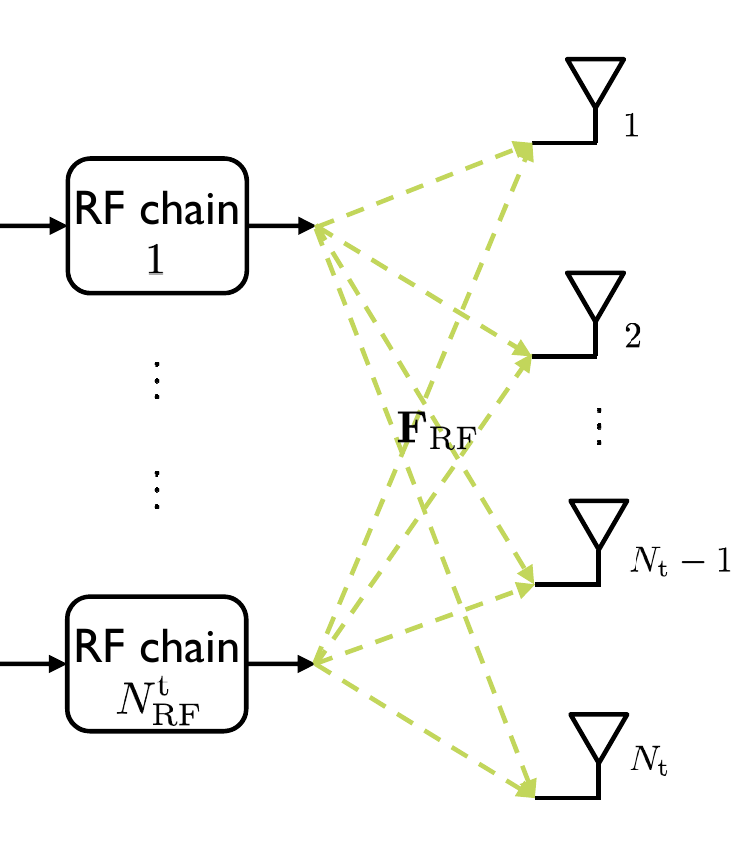}\label{s1}
	}
	\subfigure[Partially-connected mapping strategy.]{
		\includegraphics[height=5.4cm]{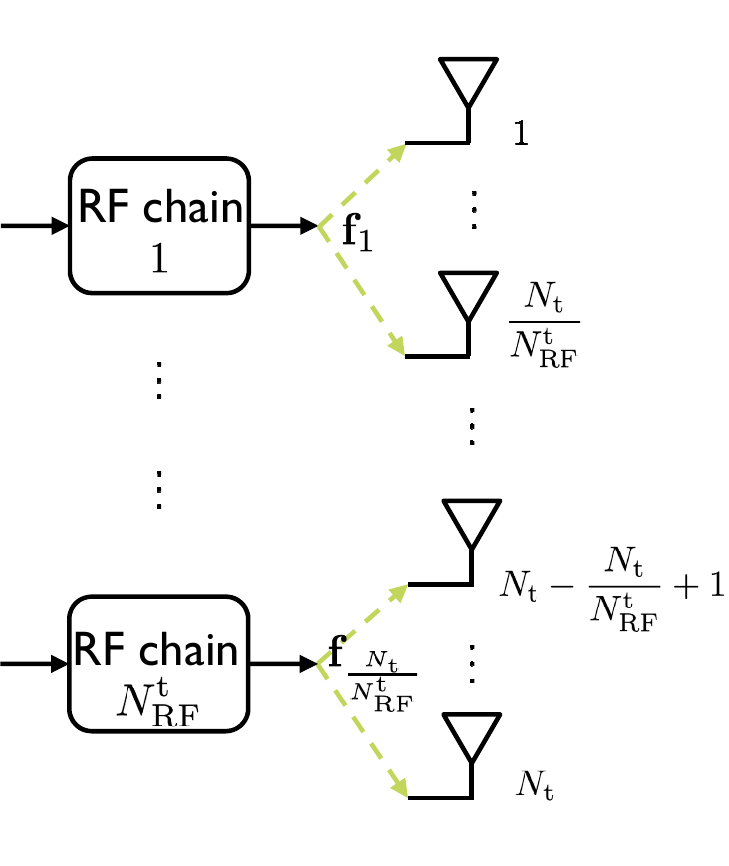}\label{s2}
	}
	\subfigure[Group-connected mapping strategy.]{
		\includegraphics[height=5.4cm]{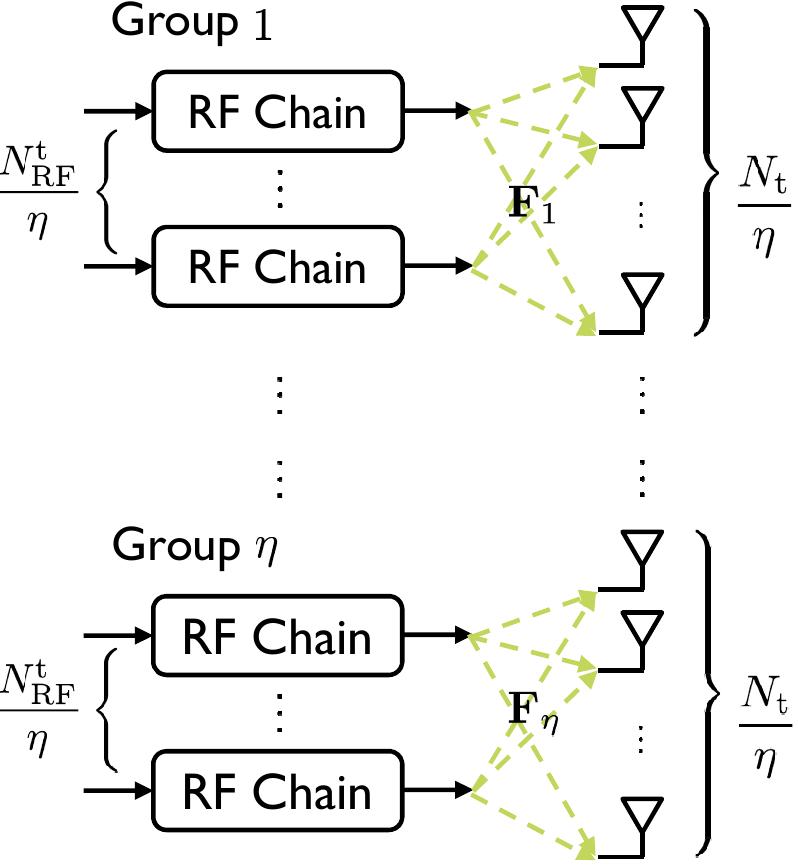}\label{s3}
	}
	\caption{Three mapping strategies for hybrid precoding in mm-wave MIMO systems: each RF chain is connected to all $N_t$ antennas in (a), to $\Nt/N_\mathrm{RF}^\mathrm{t}$ antennas in (b), and to $\Nt/\eta$ antennas in (c).}
\end{figure*}
\subsection{Hybrid Precoder Design in Multicarrier Systems}\label{SecIV}
Multicarrier techniques such as OFDM are often utilized to overcome the frequency-selective fading caused by the large available bandwidth in mm-wave systems. Compared with the narrowband hybrid precoder design in Section III, the main difference in OFDM systems is that the analog precoder is shared not only by all users but also across all subcarriers \cite{7397861,6824962}. In particular, the digital precoding matrix $\FBB\in\mathbb{C}^{\NRFt\times KN_sF}$ in $\mathcal{P}_1$ is no longer a tall matrix, since $KN_sF\ge \NRFt$ for practical OFDM system settings.

In this section, we modify the FPS-AltMin algorithm for OFDM systems. Similar to \eqref{eq5}, we enforce a semi-orthogonal constraint on the digital precoding matrix. As $\FBB$ is generally a fat matrix, the semi-orthogonal constraint is specified as
\begin{equation}
\FBB\FBB^H=\alpha ^2\FDD\FDD^H=\alpha^2\mathbf{I}_{\NRFt}.
\end{equation}
In this way, the upper bound of the objective function derived in \eqref{upper} still holds since
\begin{equation}
\begin{split}
\left\Vert\mathbf{SC}\FDD\right\Vert_F^2&=
\Tr\left(\mathbf{C}^H\mathbf{S}^H\mathbf{SC}\right)\\
&\overset{(c)}{=}\Tr\left( { {
		\begin{bmatrix}
		\mathbf{I}_{\NRFt}&\\
		&\mathbf{0}
		\end{bmatrix}	
	} {{\mathbf{M}^H}}{{\mathbf{S}^H}}{\mathbf{S}\mathbf{M}}} \right)\\
&< \Tr\left( {{{\mathbf{M}}^H}{\mathbf{S}^H}}{\mathbf{S}\mathbf{M}} \right)=\left\Vert\mathbf{S}\right\Vert_F^2,
\end{split}
\end{equation}
where (c) comes from the SVD of $\mathbf{CC}^H$, i.e., $\mathbf{CC}^H=\mathbf{M}\mathrm{blkdiag}\left(\mathbf{I}_{\NRFt},\mathbf{0}\right)\mathbf{M}^H$, since $\mathbf{C}$ is a semi-unitary matrix, and the columns of $\mathbf{M}$ are the left singular vectors of $\mathbf{CC}^H$. As the modifications in multicarrier systems lie in the digital precoding matrices $\Fopt$ and $\FBB$, in the modified AltMin algorithm, the update of $\alpha$ and $\mathbf{S}$ is the same as that in Section \ref{SecIIIB}. On the other hand, since $\FDD$ is a fat matrix in OFDM systems, the optimization of $\FDD$ should be modified as
\begin{equation}\label{eq24}
\FDD=\mathbf{V}\mathbf{U}_1^H,
\end{equation}
where $\alpha\Fopt^H\mathbf{SC}=\mathbf{U}_1\mathbf{\Sigma V}^H$ and $\mathbf{\Sigma}$ is a diagonal matrix with non-zero singular values $\sigma_1,\cdots,\sigma_{\NRFt}$, which is the SVD of $\Fopt^H\mathbf{SC}$. Correspondingly, the construction of the initial $\FDD^{(0)}$ is given by
\begin{equation}\label{eq25}
\FDD^{(0)}=\mathbf{V}^H_{\left[1:\NRFt\right]},
\end{equation}
where $\Fopt=\mathbf{U}\mathbf{\Sigma V}^H$ is the SVD of $\Fopt$ and the subscript $[1:n]$ denotes the first to the $n$-th columns of a matrix.

By substituting \eqref{eq25} and \eqref{eq24} into Steps 1 and 4 in the FPS-AltMin algorithm, respectively, we obtain the modified FPS-AltMin algorithm for mm-wave OFDM systems. The conclusion on convergence remains the same as was discussed in Section \ref{SecIIIB} while the computational complexity is $\mathcal{O}\left(KN_sF{\NRFt}^2+N_c\NRFt\Nt\log N_c\NRFt\Nt\right)$. Furthermore, the inter-user interference canceling approach can also be extended to OFDM systems, i.e., an additional BD precoder is utilized based on the effective channel that is defined as
\begin{equation}
\mathbf{\hat H}_{k,f}={\mathbf{W}^H_\mathrm{BB}}_{k,f}{\mathbf{W}^H_\mathrm{RF}}_{k}\mathbf{H}_{k,f}{\FRF}{\FBB}_{f},
\end{equation}
where ${\FBB}_f=\left[{\FBB}_{1,f},\cdots,{\FBB}_{k,f},\cdots,{\FBB}_{K,f}\right]$ with dimension ${\NRFt\times KN_s}$ is the composite digital precoder on the $f$-th subcarrier. Therefore, the extension to multicarrier systems does not lead to extra design difficulties compared with single-carrier systems.

\section{The Group-Connected Mapping Strategy for Hybrid Precoding}\label{SecV}
In previous sections, the hybrid precoder design is based on a novel hardware implementation but with a conventional mapping strategy, i.e., the fully-connected mapping. In this section, a new mapping strategy, called the \emph{group-connected} mapping, is proposed to offer a flexible trade-off between hardware complexity and spectral efficiency. In particular, with this mapping strategy, the number of switches in the FPS implementation is further reduced.

 \subsection{The Group-Connected Mapping Strategy}
 Fig. 3 compares different mapping strategies. In the group-connected mapping, the RF chains and antennas are divided into $\eta$ groups, as shown in Fig. \ref{s3}. Within each group, the mapping strategy is the same as the fully-connected
 mapping, i.e., each RF chain is connected to all $\frac{\Nt}{\eta}$ antennas. Thus, the analog precoding matrix $\FRF$ has the block diagonal structure, with each block corresponding to one RF chain-antenna group, specified as
 \begin{equation}
 \FRF=
 \begin{bmatrix}
 \mathbf{R}_1&&&\\
 &\mathbf{R}_2&&\\
 &&\ddots&\\
 &&&\mathbf{R}_{\eta}
 \end{bmatrix},
 \end{equation}
 with $\mathbf{R}_i\in\mathbb{C}^{\frac{\Nt}{\eta}\times\frac{\NRFt}{\eta}}$ being the analog precoding matrix in the $i$-th group. Note that while the RF chains and antennas are uniformly divided into $\eta$ groups in Fig. \ref{s3} to simplify notation, the grouping can be flexible, i.e., the numbers of RF chains and antennas in different groups can be different. 
 
The proposed group-connected mapping is a general mapping strategy that incorporates existing mapping strategies as special cases:
\begin{itemize}
	\item When $\eta=1$, which means that all RF chains and antennas are in the only one group, the group-connected mapping reduces to the fully-connected one, as shown in Fig. \ref{s1}.
	\item When $\eta=\NRFt$, which means there is only one RF chain in each group, and each of them is connected to $\Nt/\NRFt$ antennas, as shown in Fig. \ref{s2}, the mapping strategy corresponds to the partially-connected one, and the analog precoding matrix $\FRF$ is a block diagonal matrix with each block being an $\Nt/\NRFt$-dimension vector \cite[Eq. 29]{7397861}.
\end{itemize}

Inevitably, trade-offs need to be made among hardware complexity and spectral efficiency. The two existing mapping strategies provide such a trade-off, but in an extreme way. The fully-connected mapping strategy is with too low hardware efficiency, while the partially-connected one incurs too much performance degradation. In contrast, it will be shown later in Section \ref{SecVI} that the group-connected mapping provides a smoother transition between the two extreme cases. To the best of the authors' knowledge, this is the first proposal for a general mapping strategy in hybrid precoding systems.

Similar to existing mapping strategies, the group-connected mapping can also be applied to hybrid precoding along with any hardware implementations, e.g., SPS, DPS, and FPS implementations. As this paper mainly focuses on the FPS hardware implementation, we will elaborate the hybrid precoder design with the FPS group-connected structure in the following.

\floatname{algorithm}{FPS-AltMin Algorithm:}
\begin{algorithm}[t]
	\caption{A Low-Complexity Hybrid Precoding Algorithm for the FPS Group-Connected Structure}
	\begin{algorithmic}[1]
		\REQUIRE
		$\Fopt$
		\IF{$\frac{\NRFt}{\eta}\ge KN_sF$} 
		\STATE Construct an initial point for $\FDD^{(0)}$ according to \eqref{ini1} and set $k=0$;
		\REPEAT
		\STATE Fix $\FDD^{(k)}$, optimize $\alpha^{(k)}$ and $\mathbf{S}^{(k)}$ according to \eqref{eq20} and \eqref{eq15}, respectively;
		\STATE Fix $\mathbf{S}^{(k)}$ and $\alpha^{(k)}$, update $\FDD^{(k)}$ with \eqref{uv}; 
		\STATE $k\leftarrow k+1$;
		\UNTIL convergence.
		\ELSE 
		\STATE Construct an initial point for $\FDD^{(0)}$ according to \eqref{eq25} and set $k=0$;
		\REPEAT
		\STATE Fix $\FDD^{(k)}$, optimize $\alpha^{(k)}$ and $\mathbf{S}^{(k)}$ according to \eqref{eq20} and \eqref{eq15}, respectively;
		\STATE Fix $\mathbf{S}^{(k)}$ and $\alpha^{(k)}$, update $\FDD^{(k)}$ with \eqref{eq24}; 
		\STATE $k\leftarrow k+1$;
		\UNTIL convergence.
		\ENDIF
		\STATE Compute the additional BD precoder ${\FBD}_{k,f}$ at the baseband to cancel the inter-user interference \cite{asilomar}, and calculate the normalization factor $\kappa$ according to \eqref{kappa} for the hybrid precoder at the transmit end.
		\RETURN $\FRF=\mathbf{SC}$ and ${\mathbf{F}_\mathrm{B}}_{k,f}=\alpha\sqrt{\kappa}{\FDD}_{k,f}{\FBD}_{k,f}$.
	\end{algorithmic}
\end{algorithm}
\subsection{Hybrid Precoder Design for the FPS Group-Connected Structure}
As mentioned before, the number of RF chains and phase shifters has already been reduced by the FPS implementation. On the other hand, the amount of switches depends on the number of connections, which in turn is determined by the mapping strategy. 
For the group-connected structure, the analog precoding matrix can be rewritten as
\begin{equation}
\FRF=\mathbf{SC}=\mathrm{blkdiag}\left(\mathbf{S}_1\mathbf{\tilde{C}},\cdots,\mathbf{S}_\eta\mathbf{\tilde{C}}\right),
\end{equation}
where $\mathbf{\tilde{C}}\in\mathbb{C}^{\frac{N_c\NRFt}{\eta}\times\frac{\NRFt}{\eta}}$ is a block diagonal matrix that extracts the first $\NRFt/\eta$ blocks from the matrix $\mathbf{C}$ in \eqref{psmatrix}, and $\mathbf{S}_i$ with dimension ${\frac{\Nt}{\eta}\times \frac{N_c\NRFt}{\eta}}$ is the switch matrix for the $i$-th group.
Hence, there are ${\Nt\NRFt}/{\eta}$ RF chain-antenna pairs, and the number of switches in use is ${\Nt\NRFt N_c}/{\eta}$, which is reduced by the factor of $\eta$ compared with the FPS fully-connected structure.
Furthermore, the hardware implementation of the analog network is simplified with the group-connected mapping. In particular, with the conventional fully-connected mapping, $N_\mathrm{t}$-way power dividers and $N_\mathrm{RF}^\mathrm{t}$-way power combiners are required \cite{7928302}. In contrast, with the proposed group-connected mapping, only $N_\mathrm{t}/\eta$-way power dividers and $N_\mathrm{RF}^\mathrm{t}/\eta$-way power combiners are needed.

Fortunately, the reduced hardware complexity does not incur additional difficulties and computational complexity in hybrid precoder design. Due to the block diagonal structure of $\FRF$, the product of $\FRF$ and $\FBB$ can be expressed as
\begin{equation}
\begin{split}
\FRF\FBB&=\begin{bmatrix}
\mathbf{R}_1\mathbf{B}_1&\cdots&\mathbf{R}_\eta\mathbf{B}_\eta
\end{bmatrix}^T\\
&=\begin{bmatrix}
\mathbf{S}_1\mathbf{\tilde{C}}\mathbf{B}_1&\cdots&\mathbf{S}_\eta\mathbf{\tilde{C}}\mathbf{B}_\eta
\end{bmatrix}^T.\\
\end{split}
\end{equation}
The matrix $\mathbf{B}_i\in\mathbb{C}^{\frac{\NRFt}{\eta}\times KN_sF}$ is the sub-matrix consisting of the $(i-1)\frac{\NRFt}{\eta}+1$-th to the $i\frac{\NRFt}{\eta}$-th rows of $\FBB$. In this way, the hybrid precoder design problem can be decoupled into $\eta$ subproblems, each of which corresponds to one group, given by
\begin{equation}\mathcal{G}_i:\quad
\begin{aligned}
&\underset{\mathbf{S}_i,\mathbf{B}_i}{\mathrm{minimize}} && \left\Vert\mathbf{F}_i-\mathbf{S}_i\mathbf{\tilde{C}}\mathbf{B}_i\right\Vert_F^2\\
&\mathrm{subject\thinspace to}&&
\mathbf{S}_i\in\{0,1\}^{\frac{\Nt}{\eta}\times \frac{N_c\NRFt}{\eta}},\\
\end{aligned}
\end{equation}
where $\mathbf{F}_i\in\mathbb{C}^{\frac{\Nt}{\eta}\times KN_sF}$ is the sub-matrix that extracts the $(i-1)\frac{\Nt}{\eta}+1$-th to the $i\frac{\Nt}{\eta}$-th rows from $\Fopt$. We can observe that each subproblem $\mathcal{G}_i$ is with the same form as $\mathcal{P}_1$ with the FPS fully-connected structure. This result is also intuitively true since the mapping strategy within each group is nothing but the fully-connected one.

Following the same procedures in Sections \ref{SecIII} and \ref{SecIV}, the subproblems $\left\{\mathcal{G}_i\right\}_{i=1}^\eta$ can be solved in a parallel fashion. The only additional step is to determine whether the matrix $\mathbf{B}_i$ is a tall or fat matrix, i.e., to decide whether $\frac{\Nt}{\eta}\ge KN_sF$ or not, since they correspond to different ways to update $\FDD$ in single-carrier and multicarrier design, respectively. For the FPS group-connected structure, the computational complexity of the proposed FPS-AltMin algorithm is $\mathcal{O}\left(\frac{KN_sF{\NRFt}^2}{\eta}+\frac{N_c\NRFt\Nt}{\eta}\log \frac{N_c\NRFt\Nt}{\eta^2}\right)$.

Note that this design methodology for the group-connected mapping is applicable to any kinds of hardware implementation. This means that the algorithm design for the group-connected mapping with any hardware implementations can be realized by directly migrating the design for the fully-connected mapping, which has been investigated in abundant existing works \cite{6717211,6928432,7335586,7389996,7397861}. It also shows the benefits of introducing this group-connected mapping from the algorithmic perspective.

\section{Simulation Results}\label{SecVI}
In this section, we evaluate the performance of the proposed FPS-AltMin algorithm via simulations. Unless otherwise specified, the BS and each user are equipped with 144 and 16 antennas, respectively, while all the transceivers are equipped with uniform planar arrays. 
The phases of the $N_c$ available fixed phase shifters are uniformly separated within $[0,2\pi]$ by
	$N_c$ equal length intervals.
Four users and 128 subcarriers are assumed when considering multiuser OFDM systems. To reduce the cost and power consumption, the minimum number of RF chains is adopted according to the assumptions in Section II-A, i.e., $\NRFt=KN_s$ and $\NRFr=N_s$. The phases of the available fixed phase shifters are uniformly separated within $[0,2\pi]$ by $N_c$ equal-length intervals. 
The nominal SNR is defined as $\frac{P}{KN_sF\sigma_\mathrm{n}^2}$, and all the simulation results are averaged over 1000 channel realizations. For the fully digital precoder, the BD precoder is adopted, which is asymptotically optimal in high SNR regimes \cite{1261332}. Furthermore, the Saleh-Valenzuela model is adopted in simulations to characterize mm-wave channels \cite{6717211,7397861}, and the frequency domain channel matrix for the $f$-th subcarrier given by \cite{6884253,7397861}
\begin{equation}
\mathbf{H}_{f}=\gamma\sum_{i=0}^{N_\mathrm{cl}-1}\sum_{l=1}^{N_\mathrm{ray}}{\alpha_{il}\mathbf{a}_\mathrm{r}(\phi_{il}^\mathrm{r},\theta_{il}^\mathrm{r})\mathbf{a}^H_\mathrm{t}(\phi_{il}^\mathrm{t},\theta_{il}^\mathrm{t})}e^{-\jmath 2\pi if/F},
\end{equation}
where $\gamma=\sqrt{\frac{\Nt\Nr}{N_\mathrm{cl}N_\mathrm{ray}}}$ is the normalization factor. The numbers of clusters and rays in each cluster are represented by $N_\mathrm{cl}$ and $N_\mathrm{ray}$, respectively. The channel gain of the $l$-th ray in the $i$-th cluster is denoted as $\alpha_{il}$. Furthermore, $\mathbf{a}_\mathrm{r}(\phi_{il}^\mathrm{r},\theta_{il}^\mathrm{r})$ $\mathbf{a}_\mathrm{t}(\phi_{il}^\mathrm{t},\theta_{il}^\mathrm{t})$ represent the receive and transmit array response vectors, where $\phi_{il}^\mathrm{r}$($\phi_{il}^\mathrm{t}$) and $\theta_{il}^\mathrm{r}$($\theta_{il}^\mathrm{t}$) stand for azimuth and elevation angles of arrival and departure, respectively. While this channel model is used in the simulation, our precoder design does not depend on the channel model and is also applicable to other more general models.

\subsection{Single-User Single-Carrier (SU-SC) Systems}
\begin{figure}[t]
	\centering
	\includegraphics[height=5.5cm]{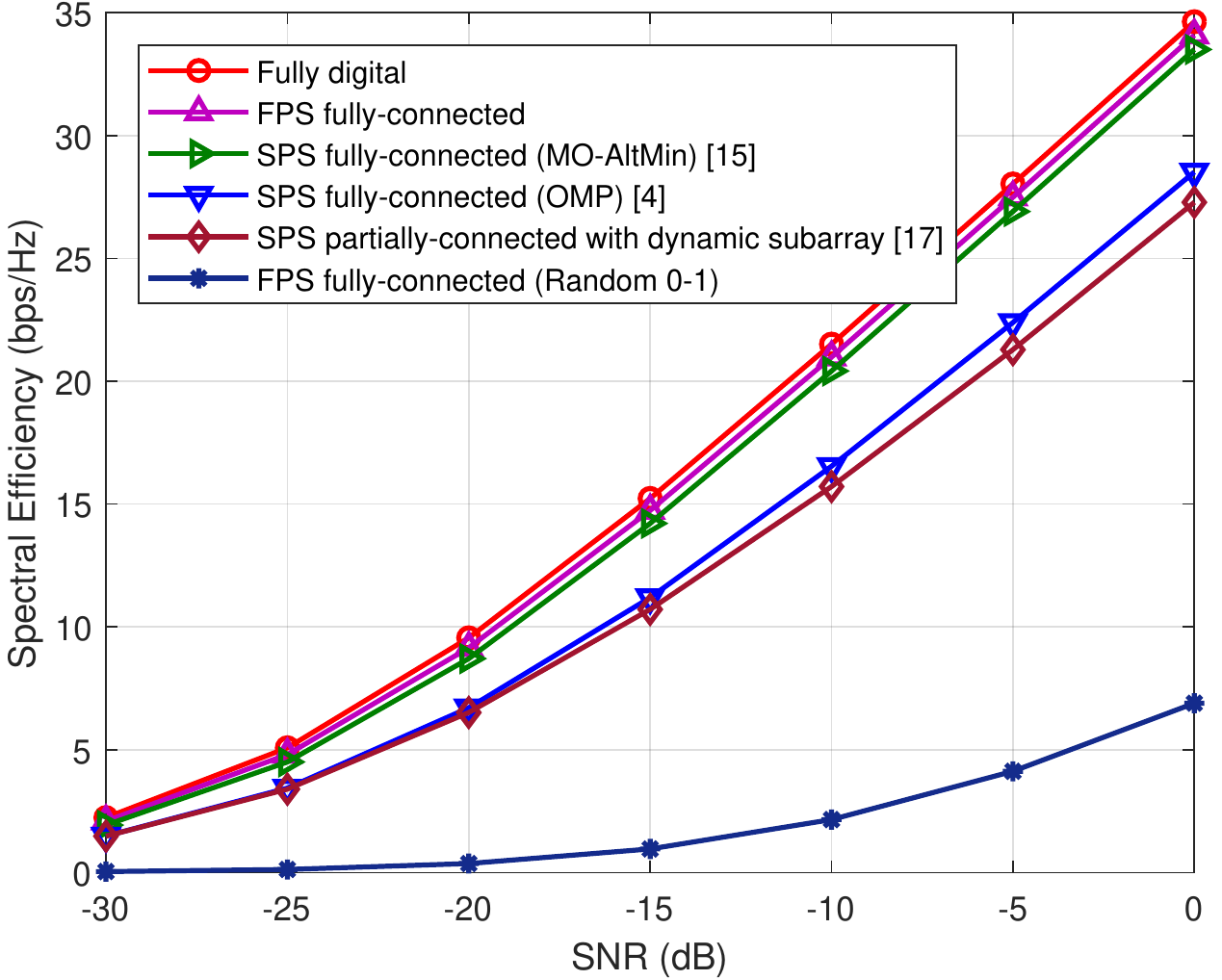}
	\caption{Spectral efficiency achieved by different hybrid precoding algorithms in SU-SC systems when $\NRFt=\NRFr=N_s=4$ and $N_c=30$.\label{fig1}
	}
\end{figure}

As a great number of previous efforts have been spent on point-to-point systems, it is intriguing to test the performance of the proposed implementation and algorithm by comparing with existing works as benchmarks.
The OMP algorithm proposed in \cite{6717211,7037444} has been widely used as a low-complexity algorithm with the analog precoder selected from a predefined set, which contains the array response vectors of the channels. An alternating minimization algorithm was then proposed in \cite{7397861} to improve the performance over the OMP algorithm, yet with high computational complexity of performing the manifold optimization, referred as the MO-AltMin algorithm. 
For the SPS partially-connected structure, a dynamic subarray approach was proposed in \cite{7880698} to compensate the performance loss caused by the fewer connections between the RF chains and antennas\footnote{As the algorithm in \cite{7880698} can only design the hybrid precoder at the BS side, a fully digital combiner is adopted at the user side for this approach while other approaches adopt hybrid combiners in Fig. \ref{fig1}.}.

In Fig. \ref{fig1}, the performance of a random binary switch matrix $\mathbf{S}$ in the FPS fully-connected structure is firstly presented. It shows that this approach is far from satisfactory and therefore a delicate design of the switch matrix is needed.
Fig. \ref{fig1} also compares the performance achieved by the proposed FPS-AltMin algorithm in the FPS fully-connected structure with three existing approaches in the SPS fully-connected structure.
It shows that, although the phase shifters are with fixed phases and the number of them is small, i.e., 30 fixed phase shifters, the proposed FPS fully-connected structure achieves the highest spectral efficiency. Thanks to the proposed low-complexity FPS-AltMin algorithm, the simulation time of the proposed algorithm is comparable to the OMP one for the SPS fully-connected structure. The performance gain in spectral efficiency over the benchmarks is mainly attributed to the proposed FPS hardware implementation, where each signal from an RF chain passes through more than one phase shifter. Furthermore, the results show that the proposed FPS-AltMin algorithm leads to an effective design of the dynamic switch network. Note that the MO-AltMin algorithm is so far the one that achieves the best performance in the SPS fully-connected structure, which means the proposed structure and algorithm stand out as an excellent candidate for hybrid precoding with high hardware efficiency, high spectral efficiency, and low-complexity design methodology.

\subsection{Multiuser Multicarrier Systems}
As we have shown that only a small number of phase shifters is required to approach the performance of the fully digital precoder in SU-SC systems, we wonder whether this phenomenon still establishes when the analog precoder is shared by all subcarriers and users in MU-MC systems.
While the MO-AltMin algorithm well tackles the unit modulus constraint induced by the SPS implementation, the extremely high computational complexity hinders its further extension to MU-MC systems where the dimension of the optimization scales up quickly.

Besides the fully digital case, we consider the following three baseline cases for comparison. A hybrid precoder design where one phase shifter is optimized in each iteration was developed in \cite{7913599}, which so far achieves the best spectral efficiency in the literature. In addition, Butler matrices can utilize fixed phase shifters and hybrid couplers to realize the SPS fully-connected structure, and the OMP algorithm is suitable for designing the analog network based on Butler matrices.
In \cite{asilomar}, the DPS fully-connected structure was proposed for MU-MC systems to approach the performance of the fully digital precoder by sacrificing the hardware efficiency of employing a large number of phase shifters, i.e., $2\Nt\NRFt$ phase shifters. In the evaluation of MU-MC systems, the DPS fully-connected structure is adopted as the benchmark, where a simple low-rank matrix approximation is enough for designing the hybrid precoder. 

\begin{figure}
	\centering
	\includegraphics[height=5.5cm]{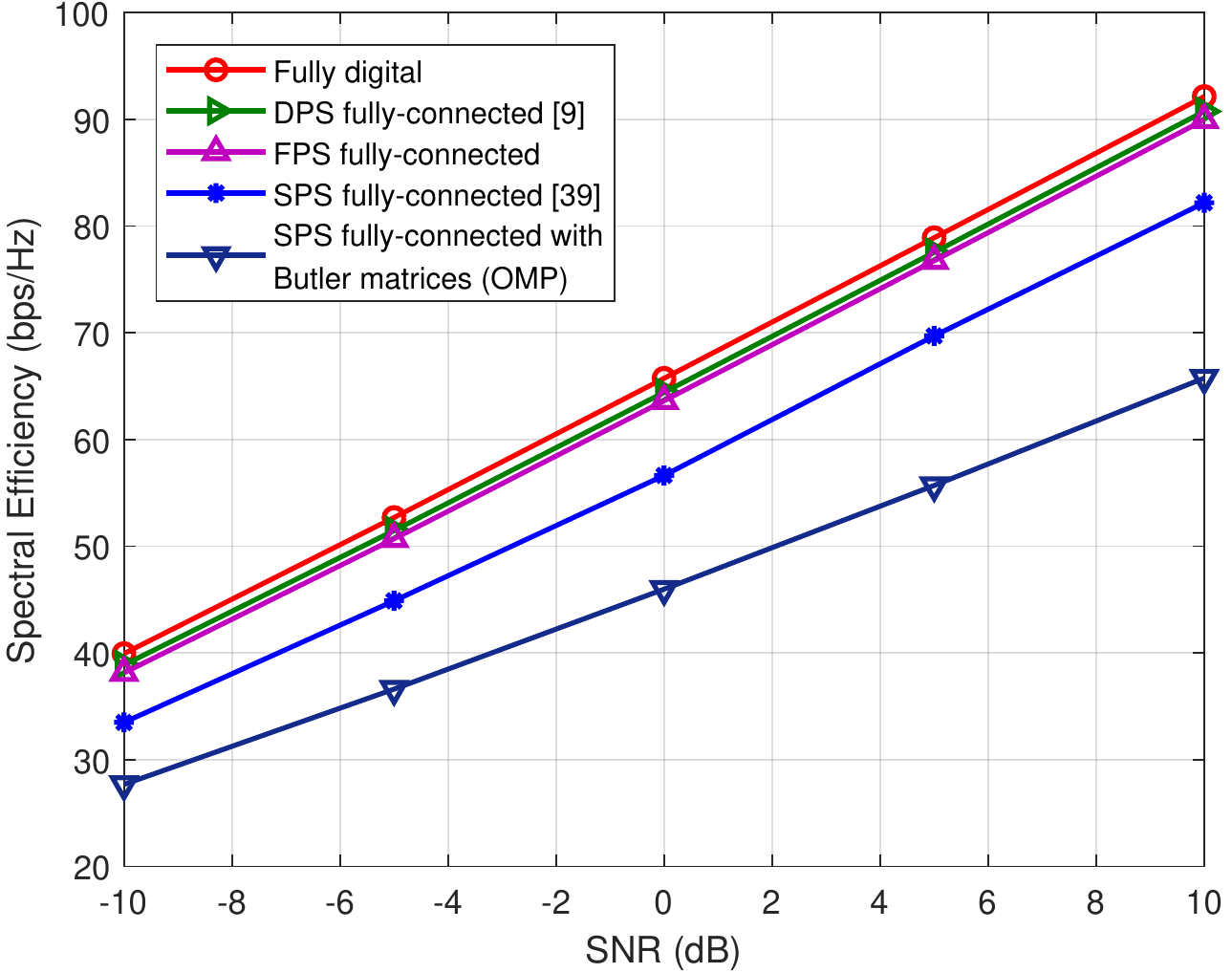}
	\caption{Spectral efficiency achieved by different hybrid precoding algorithms in MU-MC systems when $\NRFt=8$, $\NRFr=N_s=2$, and $N_c=30$.\label{fig2}
	}
\end{figure}
As shown in Fig. \ref{fig2}, the proposed FPS fully-connected structure only entails little performance loss compared to the DPS fully-connected one when only 30 fixed phase shifters are adopted. Both the DPS fully-connected and FPS fully-connected structures benefit from the operation that allows the same signal to pass through multiple phase shifters, while the main difference between them is the quantized and fixed phases assumed in the FPS one. This simulation result demonstrates that the performance loss caused by the quantization is negligible with the proposed hybrid precoder structure. 
On the other hand, the FPS fully-connected structure enjoys significant improvement in terms of spectral efficiency compared with the SPS fully-connected structure with the algorithm in \cite{7913599} and the OMP algorithm based on Butler matrices, which illustrates the effectiveness of both the newly proposed implementation and algorithm.
More importantly, it indicates that the number of phase shifters can also be sharply reduced by the proposed FPS implementation even if the analog precoder is shared in MU-MC systems. 

\begin{table*}[t]
	\begin{minipage}{0.36\linewidth}
		\centering
		\includegraphics[height=5.5cm]{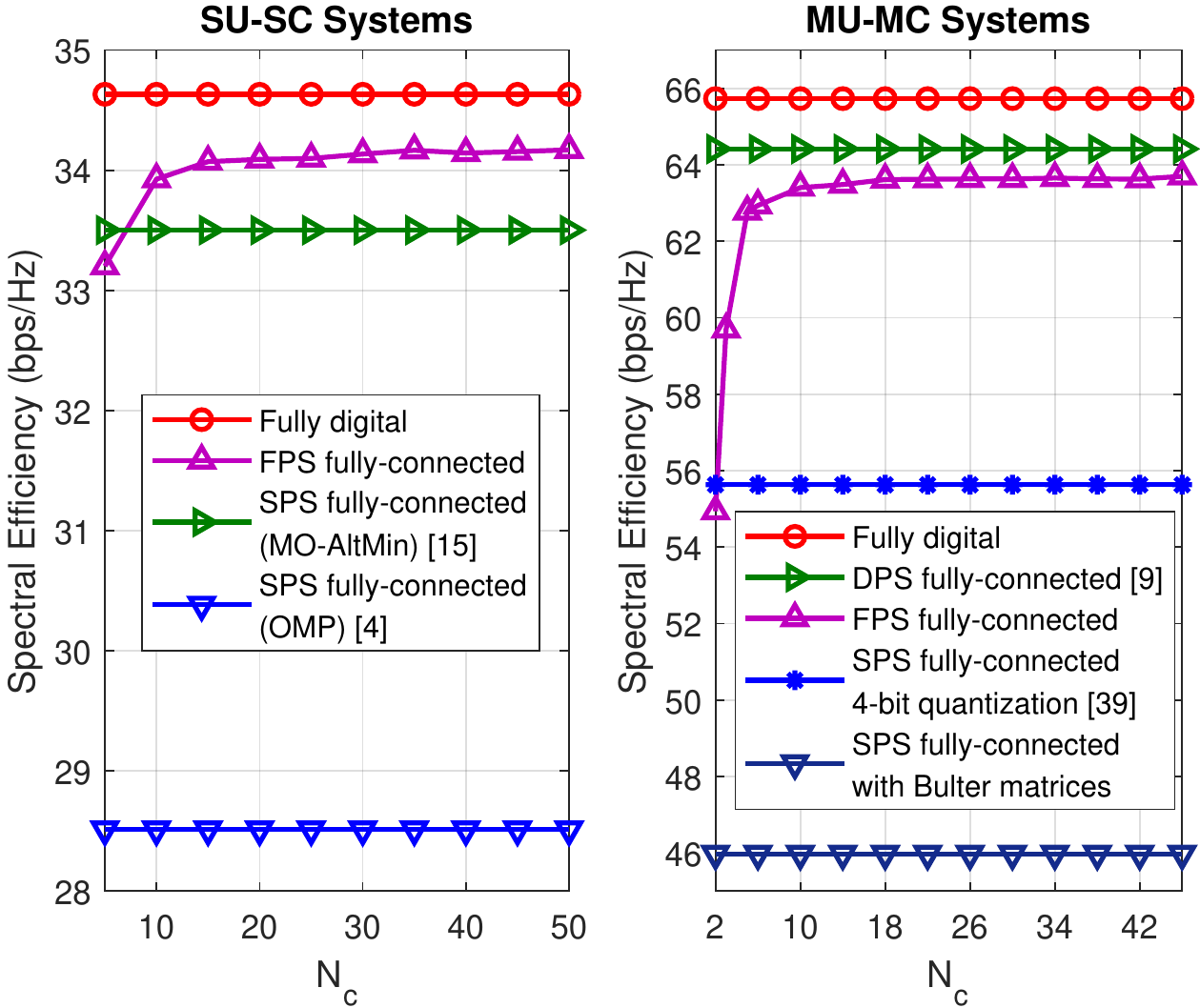}
		\captionof{figure}{Spectral efficiency achieved by different hybrid precoding algorithms in mm-wave MIMO systems given SNR$=0$ dB.}
		\label{fig3}
	\end{minipage}\hspace{0.03\linewidth}
	\begin{minipage}{0.61\linewidth}
		\centering
		\caption{Power consumption of  the analog network for different hybrid precoder structures in MU-MC systems}
		\begin{tabular}{|c|c|c|c|c|c|}
			\hline
			\multirow{2}[4]{*}{} & \multicolumn{2}{c|}{\textbf{Phase shifter}} & \multicolumn{2}{c|}{\textbf{Other hardware}} & \textbf{Total \textbf{power}\footnote{The total power consumed by the main hardware components in the analog network.}} \bigstrut\\
			\cline{2-5}          & $N_\mathrm{PS}$ & \textbf{Type} & \textbf{Hardware} & $N_\mathrm{OC}$ & $P_\mathrm{total}$ \bigstrut\\
			\hhline{|=|=|=|=|=|=|}
			\textbf{DPS fully-connected \cite{asilomar}} & 2304  & Adaptive & N/A   & N/A   & 115.2 W \bigstrut\\
			\hline
			\textbf{FPS fully-connected} & 10    & {Fixed\footnote{For fair comparisons, the power consumed by the FPS implementation is counted by calculating the power of $N_c N_\mathrm{RF}^\mathrm{t}$ fixed phase shifters, each of which is with the same power consumption as the fixed phase shifter in the Butler matrix implementation.}} & {Switch} & 11520 & 59.2 W \bigstrut\\\hhline{|=|=|=|=|=|=|}
			\textbf{SPS fully-connected} & \multirow{2}[4]{*}{1152} & \multirow{2}[4]{*}{Adaptive} & \multirow{2}[4]{*}{N/A} & \multirow{2}[4]{*}{N/A} & \multirow{2}[4]{*}{57.6 W} \bigstrut\\
			\textbf{4-bit quantization \cite{7913599}} &       &       &       &       &  \bigstrut\\\hline
			\textbf{FPS fully-connected}      & 2     &  Fixed     &   Switch    & 2304  & 11.84 W \bigstrut\\
			\hhline{|=|=|=|=|=|=|}
			\textbf{SPS fully-connected} & \multirow{2}[2]{*}{3456} & \multirow{2}[2]{*}{Fixed} & \multirow{2}[2]{*}{Coupler} & \multirow{2}[2]{*}{4032} & \multirow{2}[2]{*}{109.44 W} \bigstrut[t]\\
			\textbf{with Bulter matrices} &       &       &       &       &  \bigstrut[b]\\
			\hline
		\end{tabular}
	\end{minipage}
\end{table*}%
\subsection{Comparisons of Hardware Efficiency}
To improve the hardware efficiency, the number of fixed phase shifters, i.e., $N_c$, should be reduced to a minimum. Thus, a natural question is how many fixed phase shifters are needed to support a satisfactory spectral efficiency.
Fig. \ref{fig3} plots the spectral efficiency achieved with different numbers of fixed phase shifters, i.e., $N_c$. The simulation parameters are the same as those in Figs. \ref{fig1} and \ref{fig2} for SU-SC and MU-MC systems, respectively. Fig. \ref{fig3} shows that in SU-SC systems 15 phase shifters are enough for achieving a satisfactory performance as the spectral efficiency almost saturates when we further increase the number of fixed phase shifters. 
By contrast, 576 variable phase shifters with arbitrary precision are needed in the SPS implementation. Moreover, the OMP algorithm achieves a lower spectral efficiency and the MO-AltMin algorithm suffers from the high computational complexity. A similar phenomenon is found in MU-MC systems, i.e., around 10 fixed phase shifters are sufficient, which has not been revealed in existing works. Although the DPS implementation slightly outperforms the proposed FPS-AltMin algorithm, it employs 200 times more phase shifters with variable and high resolution. This illustrates that the proposed FPS implementation is much more hardware-efficient than existing hybrid precoder implementations, and also with satisfactory performance, which is quite attractive for practical implementation of hybrid precoding. 

As MU-MC is more likely to be the system setting in future 5G mm-wave networks, we compare the power consumption of different hybrid precoder structures in such systems, as listed in Table II. 

As the power consumption of the baseband and RF chains are the same for different hybrid precoder structures, in this section we compare the power consumption of the analog network, which is the distinct part for different structures and is mainly determined by the power consumed by phase shifters, switches or couplers. The total power consumption $P_\mathrm{total}$ of the analog network in Table II is calculated as
\begin{equation}
P_\mathrm{total}=N_\mathrm{PS}P_\mathrm{PS}+N_\mathrm{OC}P_\mathrm{OC},
\end{equation}
where $P_\mathrm{PS}$ and $P_\mathrm{OC}$ are the power consumption of each phase shifter and switch/coupler given in Table I.
For fair comparisons, we compare the hardware efficiency by calculating the power consumption of different hybrid precoder structures while keeping comparable spectral efficiency. 
As indicated in Fig. \ref{fig3}, 10 fixed phase shifters in the FPS fully-connected structure are sufficient to achieve comparable performance as that of the DPS fully-connected one. Table II shows that, while a switch network is required in the FPS fully-connected structure, it consumes much less power as the power consumption of each switch is small. This leads to a higher hardware efficiency than the DPS fully-connected structure that requires a large number of adaptive phase shifters.

On the other hand, it is found in Fig. \ref{fig3} that  2 fixed phase shifters in the FPS full-connected structure are sufficient for achieving a comparable spectral efficiency as the SPS fully-connected one with the algorithm in \cite{7913599}. Note that although infinite resolution phase shifters are assumed in \cite{7913599}, quantized phase shifters should be adopted to ensure practical comparison in terms of the power consumption. Therefore, as suggested in \cite{7370753} all the phase shifters in the SPS fully-connected structure are quantized with 4 bits. According to Table II, to achieve the same spectral efficiency, the SPS fully-connected structure needs almost 5 times more power than the FPS fully-connected one, which again demonstrates the advantages of our proposal in terms of hardware efficiency. In addition, due to the large numbers of fixed phase shifters and hybrid couplers in the Butler matrix implementation, it suffers from a huge power consumption and the lowest spectral efficiency, which results in a low hardware efficiency. Moreover, it is observed that different levels of hardware efficiency can be readily achieved by adapting the number of fixed phase shifters in the FPS fully-connected structure.

\begin{figure}[t]
	\centering
	\includegraphics[height=5.5cm]{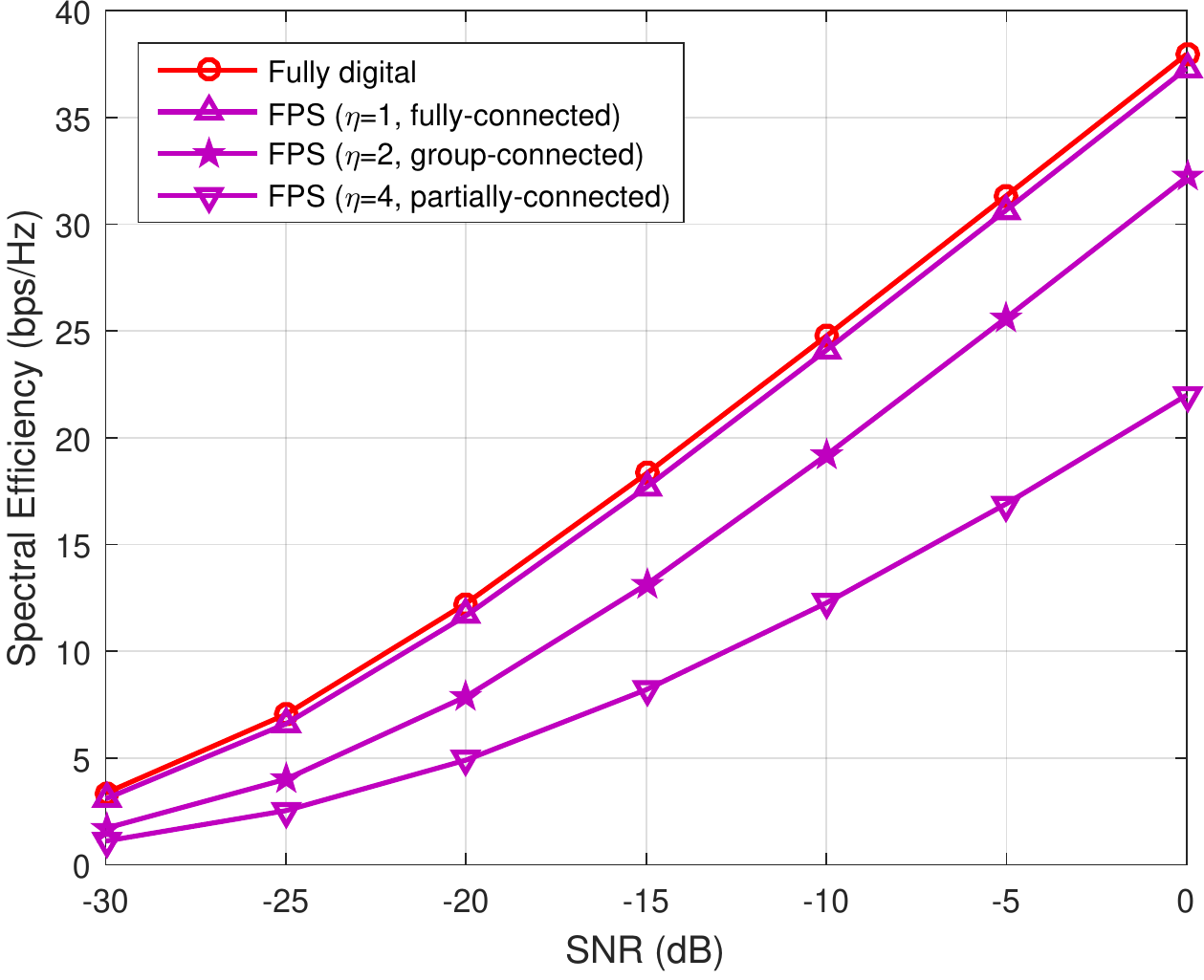}
	\caption{
		Spectral efficiency of different values of $\eta$ with the FPS group-connected structure in SU-SC systems when $\Nt=256$, $\Nr=16$, $\NRFt=\NRFr=N_s=4$, and $N_c=30$.\label{eta}
	}
\end{figure}
\subsection{The FPS Group-Connected Hybrid Precoder Structure}
\begin{figure}[t]
	\centering
	\includegraphics[height=5.5cm]{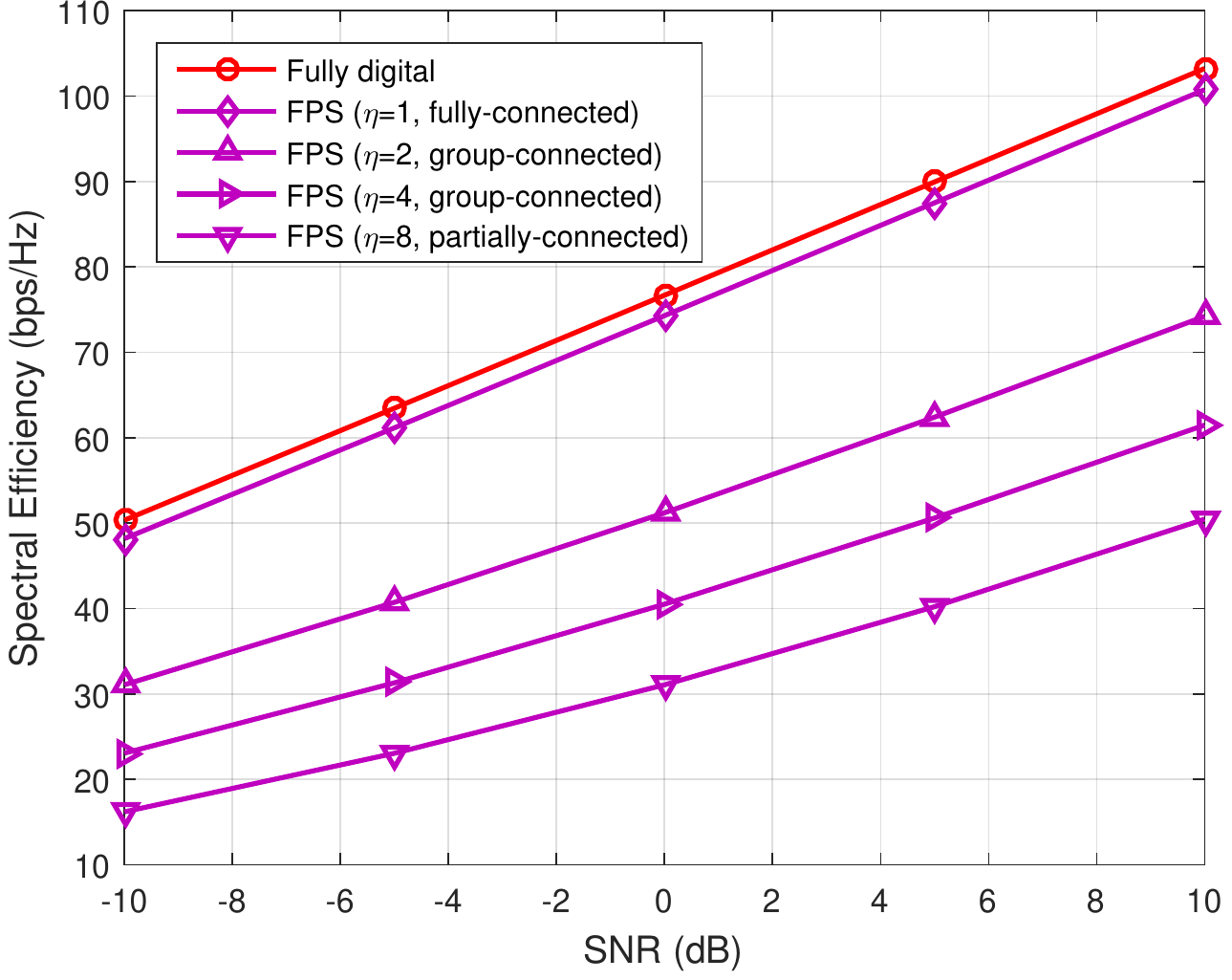}
	\caption{Spectral efficiency of different values of $\eta$ with the FPS group-connected structure in MU-MC systems when $\Nt=256$, $\Nr=16$, $\NRFt=8$, $\NRFr=N_s=2$, and $N_c=30$.\label{fig4}
	}
\end{figure}

In this part, we evaluate the spectral efficiency achieved by the proposed group-connected mapping strategy. By employing this mapping strategy with the FPS implementation, the number of switches can be reduced by a factor of $\eta$, which is the number of groups in the mapping. 
In existing works, only the fully-connected ($\eta=1$) and partially-connected ($\eta=\NRFt$) mapping strategies are available. As shown in Fig. \ref{eta}, in SU-SC systems, there is a huge gap between these two extreme mapping strategies, and the group-connected mapping provides an effective way to close this gap, which helps to balance the hardware complexity and spectral efficiency. Moreover, by varying $\eta$ from 4 to 2, the performance gap is shrunk more than a half with the number of switches being reduced by half, which shows that the superiority of the group-connected mapping in SU-SC systems. Fig. \ref{fig4} plots the spectral efficiency achieved by the group-connected mapping in MU-MC systems. Since it is more reasonable to leverage more RF chains and antennas at the BS side, we assume the group-connected structure at the BS to enable more flexible choice in $\eta$ and to show the effects of the group-connected mapping, while keep the fully-connected structure ($\eta=1$) at the user side. Since the analog precoder is a shared component, the performance gap between the group-connected mapping and the fully-connected one in MU-MC systems is enlarged when we simplify the hardware implementation of the analog precoder. Nevertheless, similar to SU-SC systems, the group-connected mapping provides a flexible approach to balance the achievable performance and hardware complexity.

\newcounter{TempEqCnt}                         			
\setcounter{TempEqCnt}{\value{equation}} 			
\setcounter{equation}{\value{longequ}}          	
\begin{figure*}
	\begin{equation}\label{longeq}
	\begin{split}
	f(\alpha)&=\left\Vert\mathbf{\tilde{x}}-\mathbf{\alpha s}\right\Vert_2^2\\
	&=\begin{dcases}
	\sum_{j=1}^i(\tilde{x}_j-\alpha)^2+\sum_{j=i+1}^{n}\tilde{x}_j^2&\alpha<0\text{ and }\frac{\alpha}{2}\in\mathcal{I}_i\\
	\sum_{j=1}^i \tilde{x}_j^2+\sum_{j=i+1}^{n}(\tilde{x}_j-\alpha)^2&\alpha>0\text{ and }\frac{\alpha}{2}\in\mathcal{I}_i\\
	\end{dcases}=\begin{dcases}
	i\alpha^2-2\sum_{j=1}^i \tilde{x}_j\alpha+\sum_{j=1}^{n}\tilde{x}_j^2&\alpha<0\text{ and }\alpha\in\mathcal{R}_i\\
	(n-i)\alpha^2-2\sum_{j=i+1}^{n}\tilde{x}_j\alpha+\sum_{j=1}^{n}\tilde{x}_j^2&\alpha>0\text{ and }\alpha\in\mathcal{R}_i\\
	\end{dcases}
	\end{split}
	\end{equation}
	\hrule 
\end{figure*}
\setcounter{equation}{\value{TempEqCnt}} 		
\section{Conclusions}\label{SecVII}
In this paper, a hardware-efficient analog network structure was developed for hybrid precoding. 
\begin{itemize}
	\item We first proposed a novel hardware implementation with a small number of fixed phase shifters, supplemented by a dynamic switch network that is adaptive to the channel states to improve the performance. The proposed FPS fully-connected structure is able to approach the performance of the fully digital precoder, remarkably, with small numbers of RF chains and phase shifters. 
\item Furthermore, a new mapping strategy for hybrid precoding was introduced. Different from existing mapping strategies that serve two extreme cases, i.e., the fully- and partially-connected mapping strategies, the proposed group-connected mapping strategy offers more refined trade-offs between hardware complexity and spectral efficiency. More importantly, this new mapping is compatible with different hardware implementations, and the hybrid precoder can be effectively designed by leveraging existing hybrid precoding algorithms.
\end{itemize}
Thus, the proposed FPS group-connected structure stands out as a promising candidate for hardware-efficient hybrid precoding in 5G mm-wave systems.
It will be interesting to consider the FPS group-connected hybrid precoder design combined with channel training and feedback \cite{8081234,7961152,7914672}, as well as to investigate the dynamic grouping for further performance improvement. 
In addition, including the matrix $\mathbf{C}$ as a design variable  to provide more theoretical support for the FPS implementation would also be a valuable future research direction.

\appendices
\section{Proof of Proposition 1}
Note that each entry in the switch matrix $\mathbf{S}$ is either $0$ or $1$, and we discover that they can be optimally determined individually once $\alpha$ is given. In particular, to minimize the objective function, $s_{m,n}$ should take value $1$ if the corresponding $(m,n)$-th entry in the matrix $\Re\left(\Fopt\FDD^H\mathbf{C}^H\right)$ is closer to $\alpha$ than $0$ in the Euclidean space, and take value $0$ otherwise, as specified in \eqref{eq15}.

The remaining problem is to choose an optimal $\alpha^\star$ that minimizes the objective function.
Since $\mathbf{S}\in\mathcal{B}$ is an element wise constraint, to simplify the notations, it is equivalent to consider the vectorization version of \eqref{eq13}, given by
\begin{equation}
\begin{aligned}
&\underset{\alpha,\mathbf{s}}{\mathrm{minimize}} && \left\Vert\mathbf{x}-\alpha\mathbf{s}\right\Vert_2^2\\
&\mathrm{subject\thinspace to}&&
\mathbf{s}\in\{0,1\}^n,
\end{aligned}
\end{equation}
where $n=\Nt N_c\NRFt$, $\mathbf{x}\triangleq\mathrm{vec}\left\{\Re\left(\Fopt\FDD^H\mathbf{C}^H\right)\right\}$, and $\mathbf{s}=[s_1,s_2,\cdots,s_n]^T\triangleq\mathrm{vec}\left\{\mathbf{S}\right\}$.

First, we sort the entries of $\mathbf{x}$ in the ascending order as $\mathbf{\tilde{x}}=[\tilde{x}_1,\tilde{x}_2,\cdots,\tilde{x}_n]^T$, where $\tilde{x}_1\le\tilde{x}_2\le\cdots\le\tilde{x}_n$. 
Then all the entries split the real line into $n+1$ intervals $\{\mathcal{I}_i\}_{i=0}^n$, where $\mathcal{I}_i\triangleq[\tilde{x}_i,\tilde{x}_{i+1}]$. Furthermore, we can obtain some insights from \eqref{eq15} to optimize $\alpha$. Specifically, if $\frac{\alpha}{2}$ falls into a certain interval $\mathcal{I}_i$, the corresponding optimal $\mathbf{s}$ can be determined as
\begin{equation}
\{s_k\}_{k=1}^{i-1}=\begin{cases}
0&\alpha>0\\
1&\alpha<0,
\end{cases}\quad
\{s_k\}_{k=i}^n=\begin{cases}
1&\alpha>0\\
0&\alpha<0.
\end{cases}
\end{equation}
\setcounter{equation}{\value{longequ}} 			
\addtocounter{equation}{1}
Therefore, the objective function in \eqref{eq13} can be rewritten as \eqref{longeq} at the top of next page.
Note that within each interval $\mathcal{R}_i=[2\tilde{x}_i,2\tilde{x}_{i+1}]$, the objective function is a quadratic function in terms of $\alpha$, and hence it is easy to give the optimal solution for $\alpha$ in Proposition 1.

\section{Proof of Lemma \ref{lem1}}
We prove Lemma 1 by contradiction. Since in each interval $\mathcal{R}_i$ the objective function is a quadratic function of $\alpha$, the optimal $\alpha^\star$ can only be obtained at the two endpoints of $\mathcal{R}_i$ or at the axis of symmetry if the objective is not monotonic in $\mathcal{R}_i$. When $\alpha<0$, the axis of symmetry of the quadratic function is given by
\begin{equation}
{\bar{x}}_i=\frac{\sum_{j=1}^i\tilde{x}_j}{i},
\end{equation}
which is the mean value of the first $i$ entries in $\mathbf{\tilde{x}}$. 

A hypothesis is firstly made that a certain endpoint $\tilde{x}_i$ is the optimal solution to $\alpha$. It means that the axis of symmetry of the objective function in $\mathcal{R}_{i-1}$ is on the right hand side of $\tilde{x}_i$, and the axis of symmetry of the objective function in $\mathcal{R}_i$ is on the left hand side of $\tilde{x}_i$, i.e.,
\begin{equation}\label{eq27}
{\bar{x}}_i<\tilde{x}_i<{\bar{x}}_{i-1}.
\end{equation}
Note that the entries in $\mathbf{\tilde{x}}$ are ordered in the ascending order. Hence, ${\bar{x}}_i$, as the mean value of the first $i$ entries in $\mathbf{\tilde{x}}$, is an increasing function with respect to $i$, i.e., ${\bar{x}}_i\ge{\bar{x}}_{i-1}$, which is contradictory with \eqref{eq27} and completes the proof for $\alpha<0$. The scenario of $\alpha>0$ can be similarly proved.

\bibliographystyle{IEEEtran}
\bibliography{bare_jrnl}

\end{document}